\documentclass{aa}
\usepackage{natbib}
\usepackage{graphicx}
\usepackage{txfonts}

\begin{document}
\title{On the eccentricity of self-gravitating circumstellar disks in 
eccentric binary systems.
}
\subtitle{}

\titlerunning{Disks in binaries}

\author{F. Marzari
        \inst{1},
        H. Scholl
        \inst{2},
        P. Th\'ebault
        \inst{3}
        and
        C. Baruteau
        \inst{4}
        }

   \offprints{F. Marzari}

   \institute{
              Dipartimento di Fisica, University of Padova, Via Marzolo 8,
              35131 Padova, Italy\\
              \email{marzari@pd.infn.it}
         \and
              Laboratoire Cassiop\'ee, Universit{\'e} de Nice Sophia Antipolis, CNRS, 
              Observatoire de la C\^ote d'Azur, B.P. 4229, F-06304 Nice Cedex, France\\
             \email{Hans.Scholl@oca.eu}
         \and
             Observatoire de Paris,
             Section de Meudon, 
             F-92195 Meudon Principal Cedex, France\\
             \email{philippe.thebault@obspm.fr}
         \and
             Astronomy and Astrophysics Department, University of California, 
             Santa Cruz, CA 95064, USA\\
             \email{clement.baruteau@ucolick.org}
             }

   \date{Received XXX ; accepted XXX}

\abstract 
{}
{We study the evolution of circumstellar massive disks around the 
primary star of a binary system focusing on the computation of disk eccentricity. 
In particular, we concentrate on its dependence on the binary eccentricity.
Self-gravity is included in our numerical simulations. 
Our standard model assumes a
semimajor axis for the binary of 30 AU, the most probable 
value according to the present binary statistics. 
}
{Two--dimensional hydrodynamical computations are performed with FARGO.
Besides the dynamical standard method to determine disk eccentricities, we
apply a morphological method which may allow a better comparison with observations.
 }
{Self--gravity leads to disks that, on average,
have low eccentricity. Moreover, the orientation of the 
disk computed with the standard dynamical method always librates
instead of circulating as in simulations without self--gravity. 
The disk eccentricity decreases with the binary eccentricity, 
a result found also in models without self--gravity.
}
{Disk self--gravity appears to be an important factor 
in determining the evolution of a massive disk in a binary system. 
High eccentric binaries are not necessarily a hostile environment for
planetary accretion.
}

\keywords{Planetary systems: formation; Planetary systems: protoplanetary disks;
          Methods: numerical}

\maketitle

\section{Introduction}

All stages of planet formation in binary star systems might be strongly
influenced by the distinctive gravitational field.
\citep{mascho00,quinta,theb04,theb06,theb08,paard08}
The protoplanetary disk around a primary star, for instance, 
is perturbed by the companion affecting its morphology and dynamical structure. 
Spiral waves develop at major resonances in the disk. Its shape is
expected to be elliptic with varying eccentricity \citep{kle8,paard08,kle82}.
Both these features
may alter the dust sedimentation process on the mid plane of the 
disk and the grain accumulation into planetesimals which might even be inhibited
(\cite{nelso}). More 
eccentric trajectories of the grains, dragged by the gas, increase the 
impact rate but, on the other hand, might lead to destructive 
collisions by increasing the relative velocity. A more turbulent 
disk might, however, favor accumulation of dust in bigger 
bodies due to gravitational instability on a faster timescale. 

Understanding the morphology of circumprimary gas disks is also crucial
for the subsequent phase of planetesimal accumulation. Indeed, several recent
studies have shown that this stage might be severely hampered, or even
stopped by the combination of secular perturbations from the companion star
and friction with gas (\cite{theb06,theb08,theb09,xizh}).
This combined effect forces a differential phasing of planetesimal
orbits according to sizes, which leads to high impact velocities for
a large fraction of planetesimal-planetesimal collisions, which could in
some cases lead to an accretion hostile dynamical environment.
\cite{paard08} have shown that these results, obtained assuming
static axisymmetric gas disks, could be further amplified
when letting the gas disk evolve and "feel" the binary perturbations.
The most accretion-hostile cases were obtained when the gaseous disk reached
high eccentricities, with impact velocities almost twice as high as in the axisymmetric case,
while low eccentricity disks gave results similar to that with a static disk
There is in principle a possibily for an eccentric gas disc to lead to lower
drag effects, that is if this eccentricity is close to the forced secular eccentricity
of the planetesimals (see Eq.17 of \cite{paard08}).
However, this theoretical behaviour is only possible for a very specific radial profile
of the gas disk density and was moreover never observed in
any test simulations, indicating that this case is probably only a marginal possibility
(see detailed discussion in Sec.6.1 of \cite{paard08}).

Also the alternative formation mechanism for giant planets 
by rapid disk gravitational instability seems to be 
affected by the companion perturbations. According to 
\cite{bo2} the presence of the secondary star might 
induce planet formation even if his results 
are questioned in \cite{maye}.

Before performing time consuming simulations of planetesimal accretion
in a binary system with a hybrid code like, for instance, 
in \cite{kleyne}, \cite{paard08} and \cite{mar_apj},
 we first investigate
the disk evolution depending on the parameters of the system.
This paper is devoted to the influence of the binary eccentricity and
self-gravity on the disk structure.

In addition,
the eccentricity of circumstellar disks in binaries may
also become in the future an observable feature. 
Next generation of interferometers (like ALMA, Atacama Large 
Millimeter/submillimeter Array)
aimed to obtain high angular
resolution millimeter and sub--millimeter images might resolve
the molecular gas and dust components of disks by deriving 
values for the disk size and ellipticity. 
\cite{lita} have already performed with VLA (Very Large Array) imaging of 
the dusty disks in the L1551 IRS 5 binary system with a resolution
of about 5 AU while with ALMA 
it might be possible to gain a factor 
from 2 to 5. The morphology of protostellar disks might be derived
by a comparison with numerical simulations. The matching of observable 
features will allow
to derive important constraints on physical disk parameters.
An important parameter for comparing observations and simulations
is the eccentricity of a disk. Classical image treatment methods developed for
measuring the eccentricity and orientation of elliptical galaxies, once 
applied to circumstellar disks may yield
different values compared to those 
usually obtained by numerical simulations. We will discuss this
problem in Section 2.

Investigating the effect of a stellar companion on the evolution of a disk surrounding
the primary, necessitates the exploration of a very large parameter space determined
in particular by
physical parameters for the disk and orbital parameters of the binary system.
In a first step, we focus on the influence of the binary eccentricity using 
for all other parameters mean standard values which are expected 
from observations. 
For the simulations, we use the latest release of the 
hydrodynamical code FARGO (\cite{mass}) which includes now in particular
disk self--gravity (\cite{baru}). Comparison with former simulations
allows the investigation of its influence on the formation of structures in the disk,
on its eccentricity and orientation. 

The paper evolves in the following way. 
We first compare in section 2 the 
standard way to derive the disk eccentricity in a simulation with image
treatment methods developed for observed disks.
In Sec. 3 we recall 
the numerical FARGO model and give the parameters of our simulated 
systems. Sec. 4  is devoted to the importance of self--gravity on the evolution of a 
disk. Sec. 5  is focused on the dependence of disk evolution on binary eccentricity.
In section 6 we summarize and discuss
our results. 

\section{Dynamical and morphological disk eccentricities}

Under the perturbations of the binary companion, the disk around the 
primary star modifies its shape. 
The three major perturbing forces acting on each gas 
particle, which are the gravitational attraction exerted by the companion star, 
the gas pressure and the gravitational attraction by the disk,
act in synergy and modify the trajectories 
of each individual gas particle. In a disk surrounding an isolated single star, 
trajectories are in a first approximation nearly circular when the self-gravitation of
the disk is neglected. The 
gravitational potential due to the disk and the companion star 
modify the trajectories which can be approximated by elliptical trajectories 
with varying eccentricities. Hydrodynamical simulations of a disk yield for each cell
of the grid the velocity of the gas which can be used to compute an 
individual eccentricity attributed to each cell. The mass-weighted average of
the eccentricities of all cells has been used as an estimator for the 
eccentricity of the disk (\cite{kle8},\cite{pine}, \cite{paard08}). 
We call it in the following the dynamical disk eccentricity
$e_d$. It is defined by:

\begin{equation}
e_d = \frac{ \sum_i e_i m_i} {M_d}
\end{equation}

where $e_i$ is the eccentricity of each cell $i$ of the grid,
assuming that the local position and velocity vectors uniquely define a
2--body Keplerian orbit, 
while $m_i$ is its mass computed from the local disk density.
$M_d$ is the disk mass. 
In a similar way, a mass averaged 
perihelion of the disk $\varpi_d$ with respect to the orientation of the binary's
apsidal line
can be computed.  
We call these the {\it dynamical} elliptical 
elements of the disk.  

If the self--gravity of the disk is included in the model, 
the calculation of $e_i$ and $\varpi_i$ must be improved.
When a two--body Keplerian orbit is derived for each gas cell
it must account for the gravitational attraction of all 
the other disk components. An approximate perturbative model 
can be used in this case as in 
\cite{mar_apj}. 
The disk is approximated as 
a sequence of uniform density rings starting from the inner radius and outwards
until the outer border of the disk is reached.
The density of each ring is calculated by averaging the local disk density 
within the ring derived from the numerical simulation. 
The gravitational attraction of each ring on each individual 
gas cell can be
analytically computed (\cite{kro}) and, being a radial force, it can 
be added to the central force of the star. A Keplerian orbit 
for each gas cell is computed by adopting a slightly increased 
mass for the star which must be updated at each timestep 
since the mass distribution within the disk changes 
because of spiral waves and disk eccentricity. 

By comparing the computation of 
the orbital elements of each disk component with and without 
the above mentioned algorithm to account for self--gravity we 
find that for eccentricities larger than 0.1 the discrepancy 
between a rough 2--body estimate of $e_i$ and $\varpi_i$ derived 
with a constant stellar mass and that obtained with a varying 
stellar mass because of the disk 
self--gravity are not very different; so we neglect this effect 
in this paper. 
We can also safely neglect the effects of pressure forces in a 
first approximation since they do not introduce a large error 
in the computation of a reliable value of both $e_i$ and $\varpi_i$. 
In conclusion, hereinafter we compute the eccentricity $e_i$ of each disk component 
by using a 2--body approximation with fixed stellar mass.

\begin{figure}
\resizebox{\hsize}{!}{\includegraphics{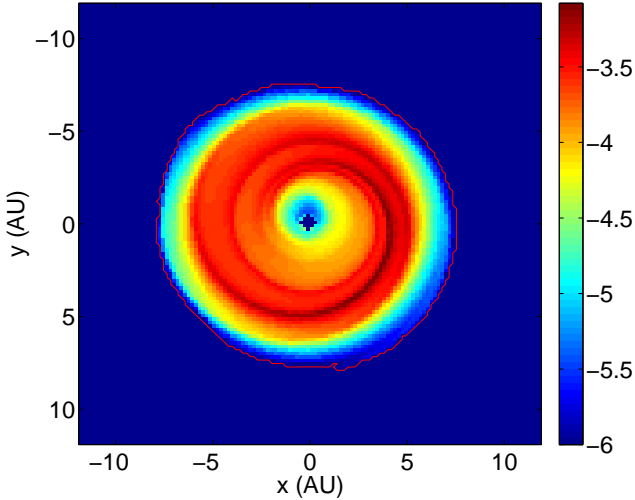}}
\caption[]{Example of outer disk contour identification
by MATLAB. The value of $\rho_0$ is set to $10^{-6}$. 
An ellipse is then fitted to the contour to compute 
$e_m$, $\varpi_m$ and $a_m$. The color scaling 
gives the 
logarithm
of the gas density in normalized units.
}
\label{f1}
\end{figure}

An important question is how to relate $e_d$ to the disk 
eccentricity an observer would obtain applying an image treatment method.
Eccentricities of elliptically shaped
objects, in particular of elliptical galaxies,
on a digitized image are obtained basically by two classical methods. 
One method uses all pixels
which form an object. Computing first and second order momenta of the object 
yields eccentricity,
semimajor axis and orientation of an ellipse approximating the object. The second method
uses contour lines formed by pixels with about the same intensity. 
An ellipse is then fitted to the contour line.
We applied the second method, which is implemented in MATLAB, where we 
force the focus of the ellipse to lay on the star. This method yields the 
 $\it morphological$ parameters $e_m$, $\varpi_m$ and $a_m$ of a disk. Fig.\ref{f1}
illustrates the morphological method.
Using the gas densities at each cell in a simulation, we produce a synthetic CCD
image. The pixel values correspond to gas densities. 
The morphological parameters depend on the choice of the intensity values of 
pixels and, hence, of a gas density value $\rho_0$ forming the contour line which
corresponds to the border of the disk.

\begin{figure}
\resizebox{\hsize}{!}{\includegraphics{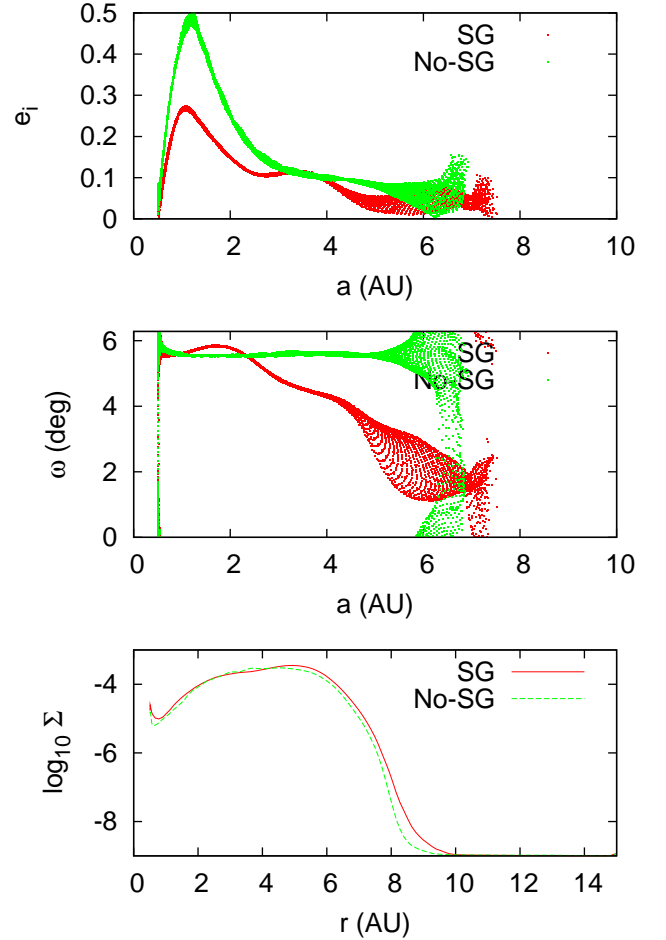}}
\caption[]{Distribution of $e_i$ (top plot) and $\varpi_i$ 
(middle plot) over the 
disk as a function of $a_i$ for both the SG and 
non--SG cases. In the bottom plot the density distribution 
averaged over the sectors is shown.
}
\label{tre}
\end{figure}

The advantages of using morphological parameters are the following:
\begin{itemize}
\item They are independent of the internal structure
of the disk
\item They are close to observable quantities 
\item The morphological semimajor axis  $a_m$ can be
used as a reliable estimate of the disk size. 
\end{itemize}
It is noteworthy that the value $e_m$ may significantly differ from 
the average value of the $e_i$'s computed for $a = a_m$  
because at the outer border of the disk the distribution 
of the $e_i$'s and $\varpi_i$'s is sparse. In Fig.\ref{tre} we show the 
distribution of both $e_i$ and $\varpi_i$ as a function of the 
cell semimajor axis $a_i$. Close to the star the values of 
$\varpi_i$ are very similar
independently of the azimuthal 
angle. However, at the outer border both the distribution open up and 
the values $e_i$ and $\varpi_i$ depend on the azimuthal angle. 
In the bottom plot we show a case in which the values of $\varpi_i$ 
are distributed in the range $[0,2\pi]$. 
For this reason  $e_m$ may be significantly different from $<e_i (a_m)>$.

The morphological method to compute disk eccentricities depends on the choice of
$\rho_0$ which is not an objective choice but very personal. This is, however, 
a very common problem for observers applying image treatment methods. 
In addition, in our simulations 
we find that the border of the disk is always well defined and it drops to 
small values very quickly due to the tidal perturbations of the 
star. This is shown in  Fig.\ref{dens} where at about 8 AU from the 
central star the disk density rapidly drops.
This makes us confident that the method to 
compute the 
overall disk eccentricity $e_m$ and its orientation $\varpi_m$
is  not very sensitive to the minimum density value 
selected in the computations when this is below $10^{-6}$ 
(obtained after exploring a significant numer of test cases)
which will be the value of $\rho_0$ always adopted in 
this paper. A test computation of $e_m$ for different values 
of $\rho_0$ ranging from $10^{-5}$ to $10^{-7}$ gave a maximum 
variation of less than 7\%.

The drastic truncation of the disk 
(see Fig.\ref{dens}) makes us also confident
that the shape derived from the morphological analysis might
be retrieved in the future from observations even if the 
disk is optically thick. The shape of an optically thin 
disk can be retrieved by observations in the millimeter 
range and the column density (our model superficial density)
is indeed proportional to the flux. For optically thick disks
this is not necessarely true, however the sharp truncation of 
the disk might be retrieved by inspecting the scattered light 
from the star. 

\section{Numerical set up}

\subsection{The hydrodynamical model}

To model the evolution of the disk we use the hydrodynamical
code FARGO \citep{mass} solving the Navier-Stokes and
continuity equations on a 2--dimensional polar grid. 
The mesh center lies at the primary and the indirect
terms are included in
the potential calculation. Since we are not interested in the 
evolution of the binary star system but only in the structure of 
the disk under the binary perturbation after the tidal truncation of the
disk, we fix the orbital parameters
of the companion star.  
In the latest release of FARGO a
Poisson equation solver based on the Fourier method
has been implemented (\cite{baru}) allowing to properly model the
disk self--gravity (hereinafter SG).
The impact of SG on the disk evolution is relevant when modelling
highly perturbed massive disks.

In all the simulations we use a grid extending from 0.5 to 15 
AU from the primary star: the number of rings is 256 while 
that of the sectors is 512. 
The aspect ratio $ h = H/r$ is constant all over
the disk and set to 0.05. The initial density profile 
declines as $\Sigma = \Sigma_0 r^{-1/2}$ and is smoothly Gaussian truncated at the inner
and outer borders. At the inner border $\Sigma$ is reduced by 
a factor 10 in betwen 0.55 and 0.5 AU, while at the outer border of the disk, 
starting from 11 AU, the density is decreased  to
$\Sigma_f = 10^{-9} \Sigma_0$ within 2 AU. 
Following \cite{kle8}, to prevent 
numerical instabilities at the outer border of the disk caused 
by very low density values, we introduce a density 
floor equal to $\Sigma_f$. Whenever the evolution 
leads to a density value lower than $\Sigma_f$,  
 $\Sigma$ is reset to $\Sigma_f$. In a series of tests this 
has proven to be a good value preventing instabilities and 
not leading to an artificial mass growth of the disk. 

We adopt a density 
at 1 AU equal to 2.5e-4 in normalized units giving an initial mass
of the disk equal to $0.04 M_{sun}$. Our disk is more massive than 
that studied in \cite{kle82} whose total mass was about 
$0.0015 M_{sun}$. Since the results do not scale only with the size
of the system but also with the mass of the disk, we expect that the 
results can be different.  

\begin{figure}[hpt]
\resizebox{\hsize}{!}{\includegraphics[angle=-90]{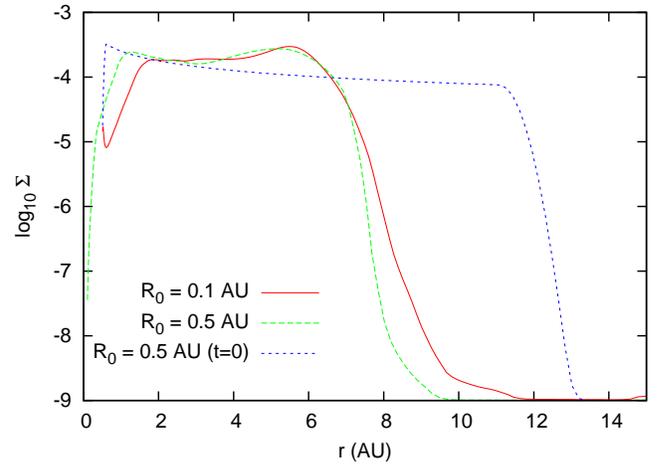}}
\caption[]{Density distribution for the simulation with the 
inned radius $R_0 = 0.1 AU$ after 4 binary orbits
compared to that with $R_0 = 0.5 AU$ at the same time. The distributions are
similar.  We include also the initial density distribution for 
$R_0 = 0.5 AU$.
}
\label{dens}
\end{figure}

 We used a non-reflecting boundary condition in all our
calculations. It is aimed at removing as many reflections as possible off the grid boundaries,
while allowing mass to flow through the inner and outer edges.
\cite{kle82} used a different kind of boundary condition to model the 
disk in $\gamma$ Cephei forcing a viscous inwards--flow of a disk at
equilibrium at the inner border. With the non--reflecting boundary
condition we have a mass inwards--flow which is comparable to that 
computed by \cite{kle82} in terms of mass fraction of the disk 
and we do not observe any strong outflow 
of disk material during the periastron phase through the inner border.
A small elliptic central hole in the disk develops with time (see
Fig.\ref{f1}), a 
consequence of the eccentric orbits
of the gas in the proximity of the star. 
Since our grid extends down to 0.5 AU, all 
gas particles with pericenter $q$ lower than 0.5 AU are expected 
to exit the inner border of the disk. Due to the collimated values 
of the perihelia $\varpi$ the outflow of the particles with 
$q < 0.5$ AU creates a region of low density with an elliptical shape
centered on the star (see Appendix A). 
If the $\varpi_i$'s were random, then the inner hole would have been 
circular with a radius larger than 0.5. In support of this interpretation 
we find that the size of the elliptic hole in the simulations is well 
reproduced by modelling its outer shape with
a Keplerian orbit with the pericenter at 0.5 AU and 
with eccentricity equal to the average eccentricity $e_i$ of the local gas
cells. 
In conclusion, with our
boundary condition, the code produces
reasonable
results taking into account the truncation of the disk at the 
inner edge. 
Of course, setting the inner rim
of the disk to 0.5 AU is an overestimate since most observative models
assume that disk truncation occurs near corotation with the inner
region cleared by the magnetosphere (\cite{shu}). In this context, 
the inner elliptic hole in our simulations can be considered a numerical artifact
since fluid elements having their pericenter at
the inner edge, which is larger than the physical one,
are not allowed to return into the grid. 
To test the influence of this issue on the disk eccentricity we 
performed a test simulation of our standard case with the 
inner border at 0.1 AU. This should be a 
realistic simulation even if it appears impossible to get it
far in time due to the short timestep imposed by the CFL condition. 
After 4 binary orbits the disk eccentricities in the two simulations 
differ by less than 5 \% and the two density distributions, 
shown in Fig.\ref{dens} appear similar. This makes us confident
that, in spite of the larger inner hole produced in the simulations
with the inner border at 0.5 AU, the models are still reliable
in terms of disk eccentricity computation. 

\subsection{Our standard binary system}

We select our representative binary system by inspecting the orbital 
element distributions of binary systems given in \cite{duma}. From their
histograms we derive the most probable orbital elements and mass ratio
for most wide binaries in our neighborhood. The semimajor axis is chosen to be 
$a_b = 30 AU$ and  the eccentricity $e_b = 0.4$. The mass of the primary
star is set to $M_p = 1 M_{\odot}$ while that of the secondary is 
$M_s = 0.4 M_{\odot}$. The orbital period of the companion star is 
about 134 yrs. This should represent the most frequent configuration
where planet formation may occur in binary systems. At the same time it is 
a very good example of a highly perturbing configuration due to the 
large eccentricity and mass of the binary. We adopt in most of 
our simulations a value of kinematic
viscosity of $10^{-5}$ (normalized units) 
which corresponds, at about 5 AU of the disk, 
to an $\alpha$ value of about $2.5 \times 10^{-3}$ (\cite{shak}). 

To study the dependence of the disk eccentricity on $e_b$ we have varied 
this parameter from 0.0 to 0.6 at a constant step of 0.1. We have also 
run a case without viscosity (inviscid case) to compare with the 
viscous case. 
 
\section{Effect of self--gravity}

\begin{figure}
\begin{center}
\resizebox{\hsize}{!}{\includegraphics[angle=-90]{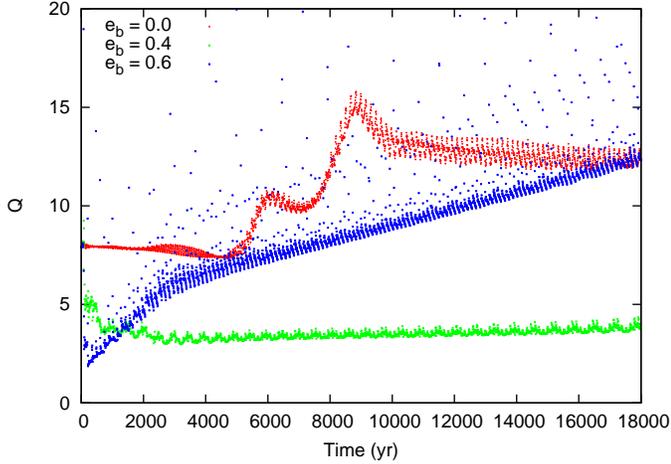}} 
\caption[]{Average value of the Toomre parameter Q computed at $r = 4 $AU from 
the primary star for different models 
with $e_b = 0$ (red dots), $e_b = 0.4$ (green dots) and $e_b = 0.6$ 
(blue dots). This last case shows large variations due to the strong 
disk perturbations when the companion star is at periapsis.
}
\label{tom}
\end{center}
\end{figure}

In our simulations we include the effect of self--gravity. The typical 
parameter adopted to measure the relevance of self--gravity is 
the Toomre parameter $Q$:
\begin{equation}
Q = \frac {h M_{p}} {\pi r^2 \Sigma}
\end {equation} 
where $r$ is the radial distance from the primary star,
whose mass is $M_p$, and 
$\Sigma$ is the surface density of the disk. 
This formulation holds for our models where the disk is isothermal 
and it has a constant aspect ratio $h = H/r$. In 
Fig.\ref{tom} we show the value of $Q$ averaged over 
several rings  around 4 AU
for three simulations 
with different binary eccentricity ($e_b = 0, 0.4, 0.6$, respectively). 
In all cases $Q$ is below 15 indicating that indeed the disk gravity may
play a role in shaping the disk. In addition, as suggested by 
\cite{paard08}, the parameter that might really measure the relevance of 
self--gravity is $ h \cdot Q$ which is well below 1 in our cases. The large fluctuations
of $Q$ observed in the case with $e_b = 0.6$ are due to the larger perturbations 
of the companion star when it is at the pericenter of a highly eccentric orbit. 

\begin{figure*}[hp]
\begin{center}
\begin{tabular}{c c}
\resizebox{90mm}{!}{\includegraphics[angle=-90]{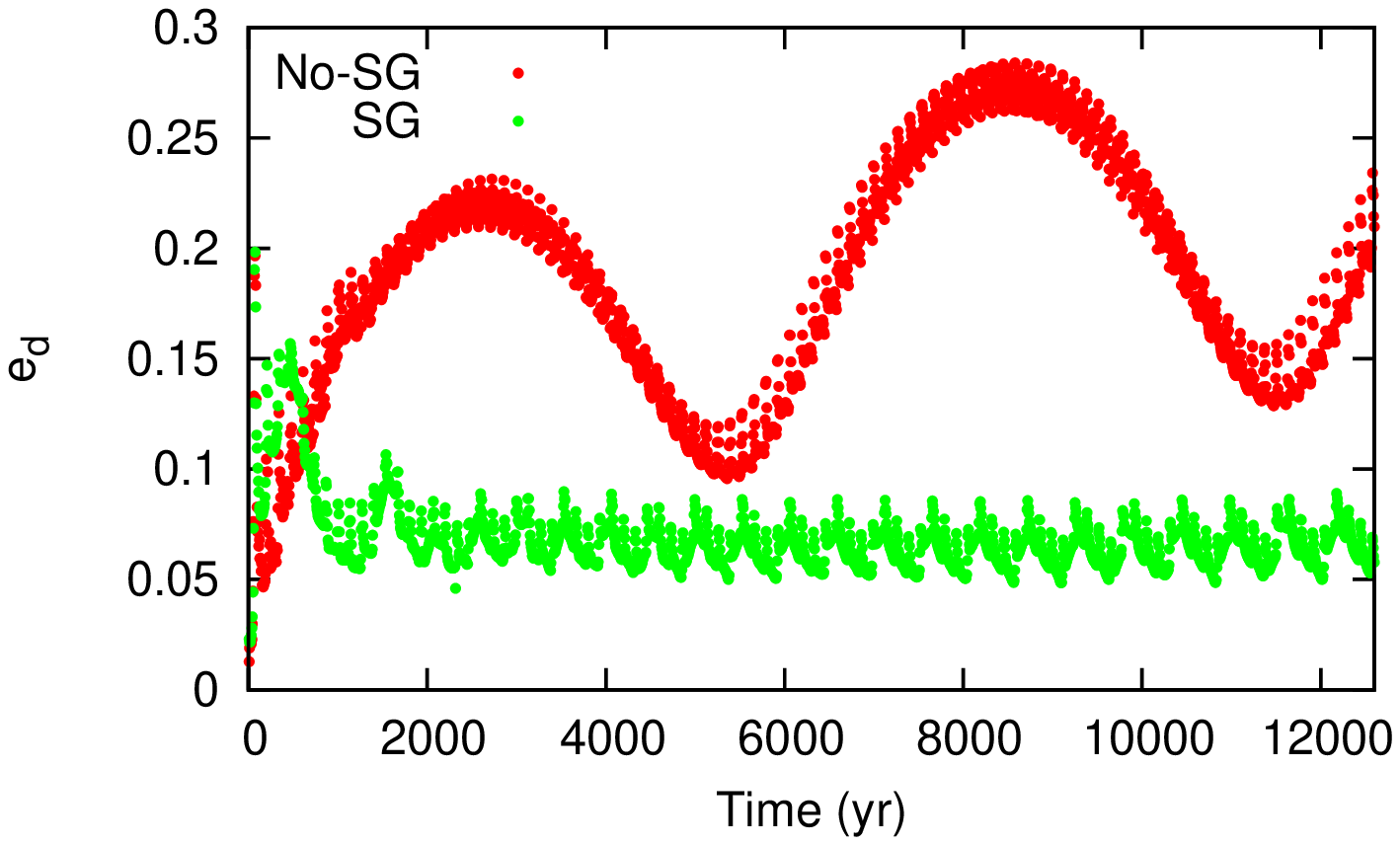}} 
\resizebox{90mm}{!}{\includegraphics[angle=-90]{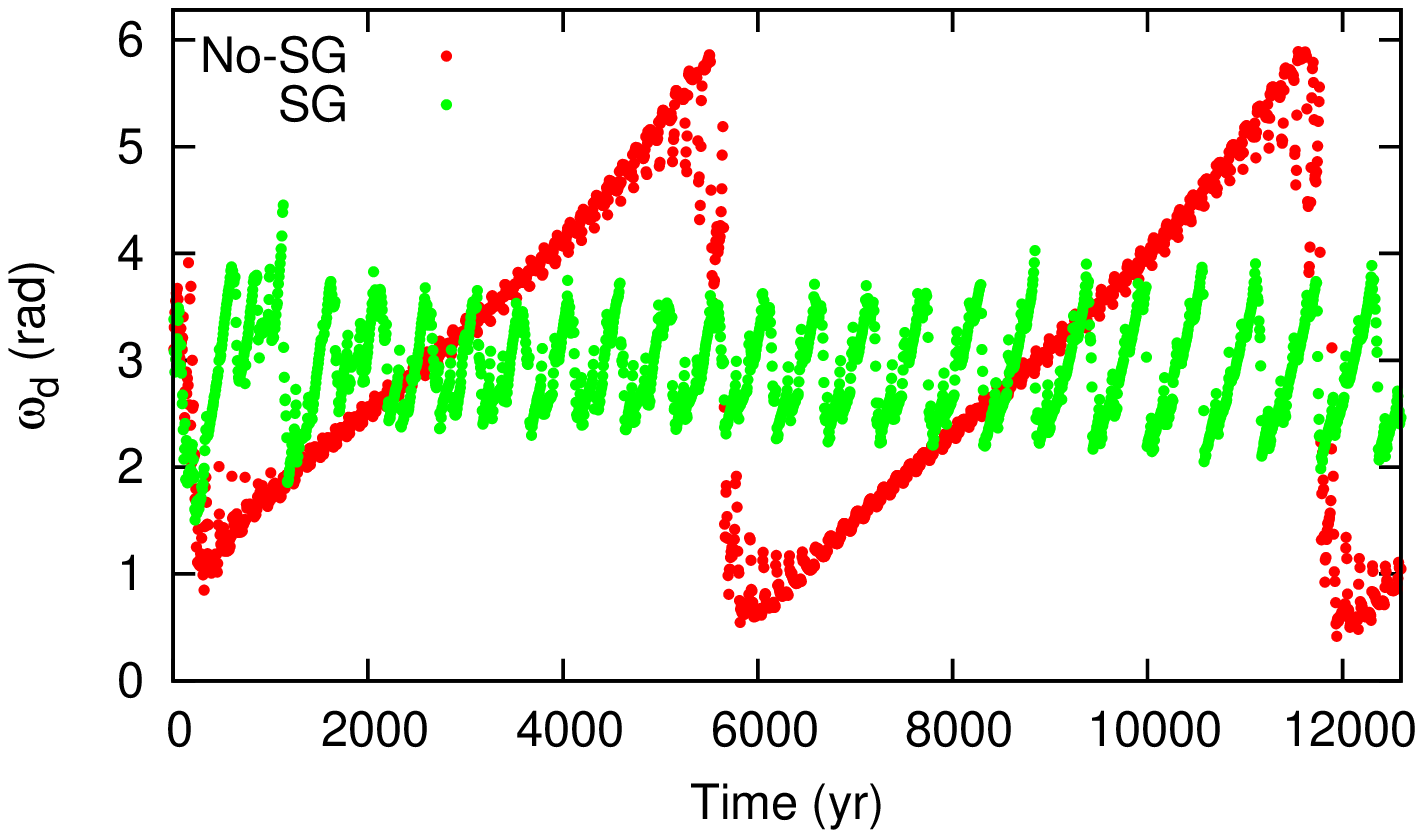}} \\
\resizebox{90mm}{!}{\includegraphics[angle=-90]{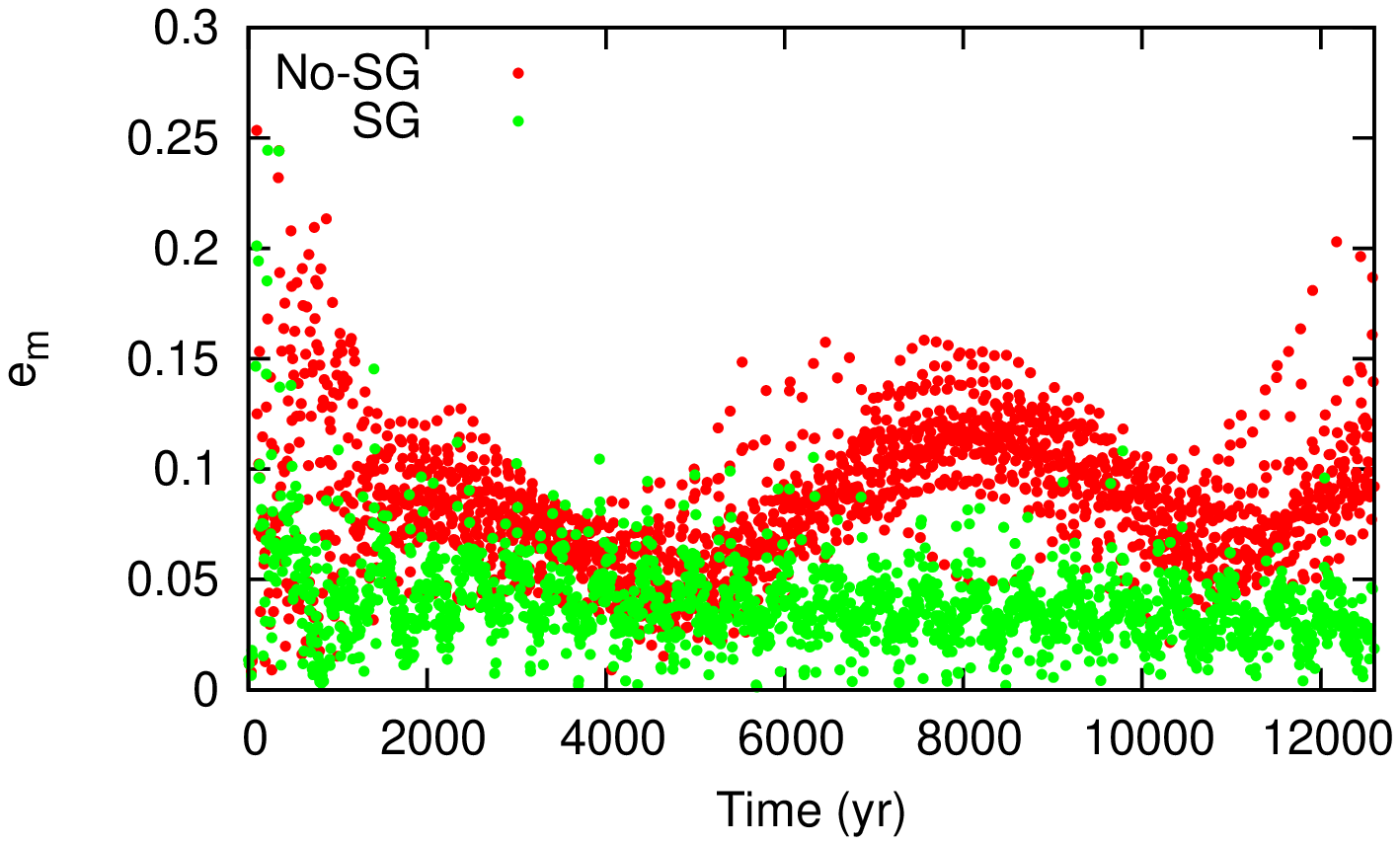}} 
\resizebox{90mm}{!}{\includegraphics[angle=-90]{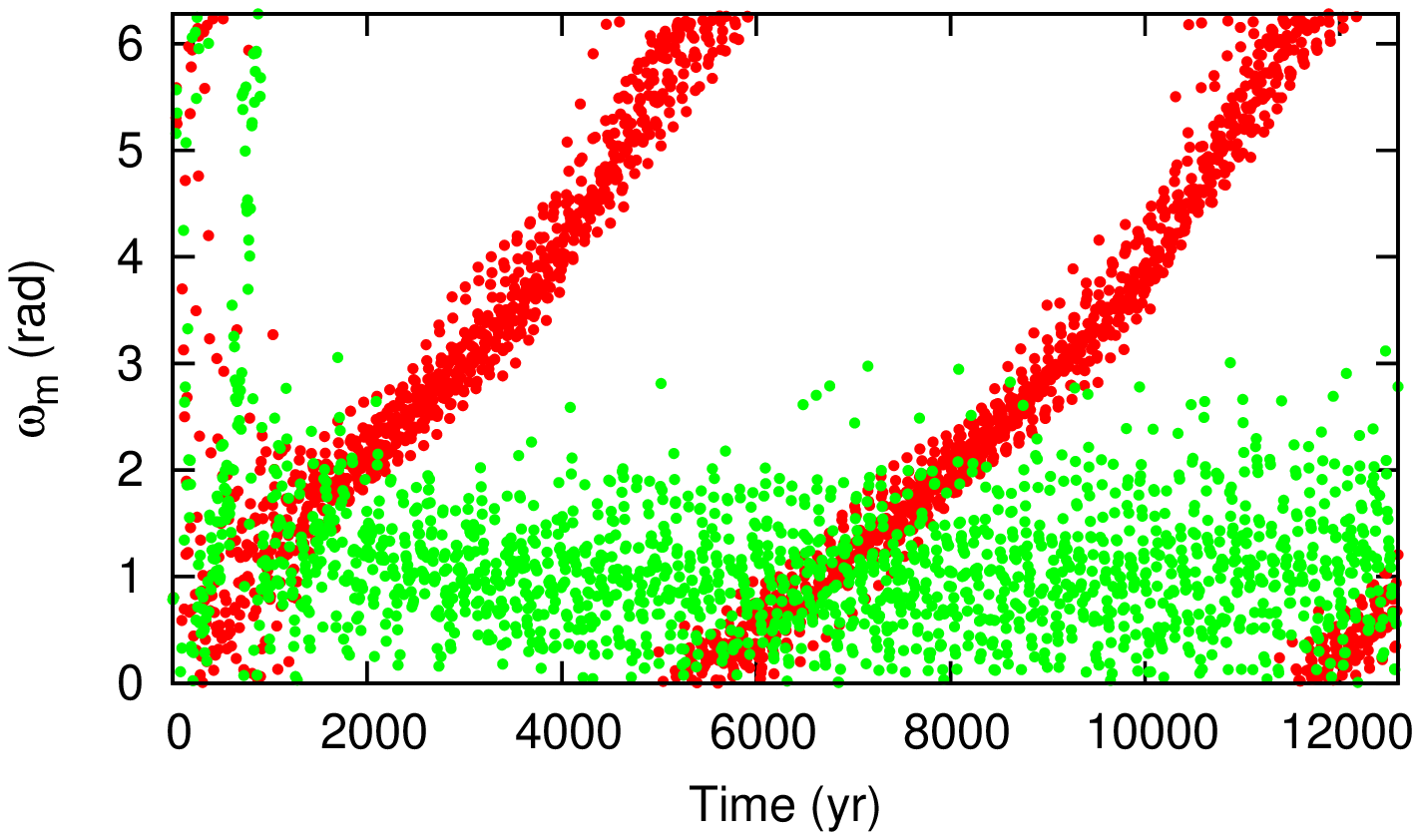}} \\
\resizebox{90mm}{!}{\includegraphics[angle=-90]{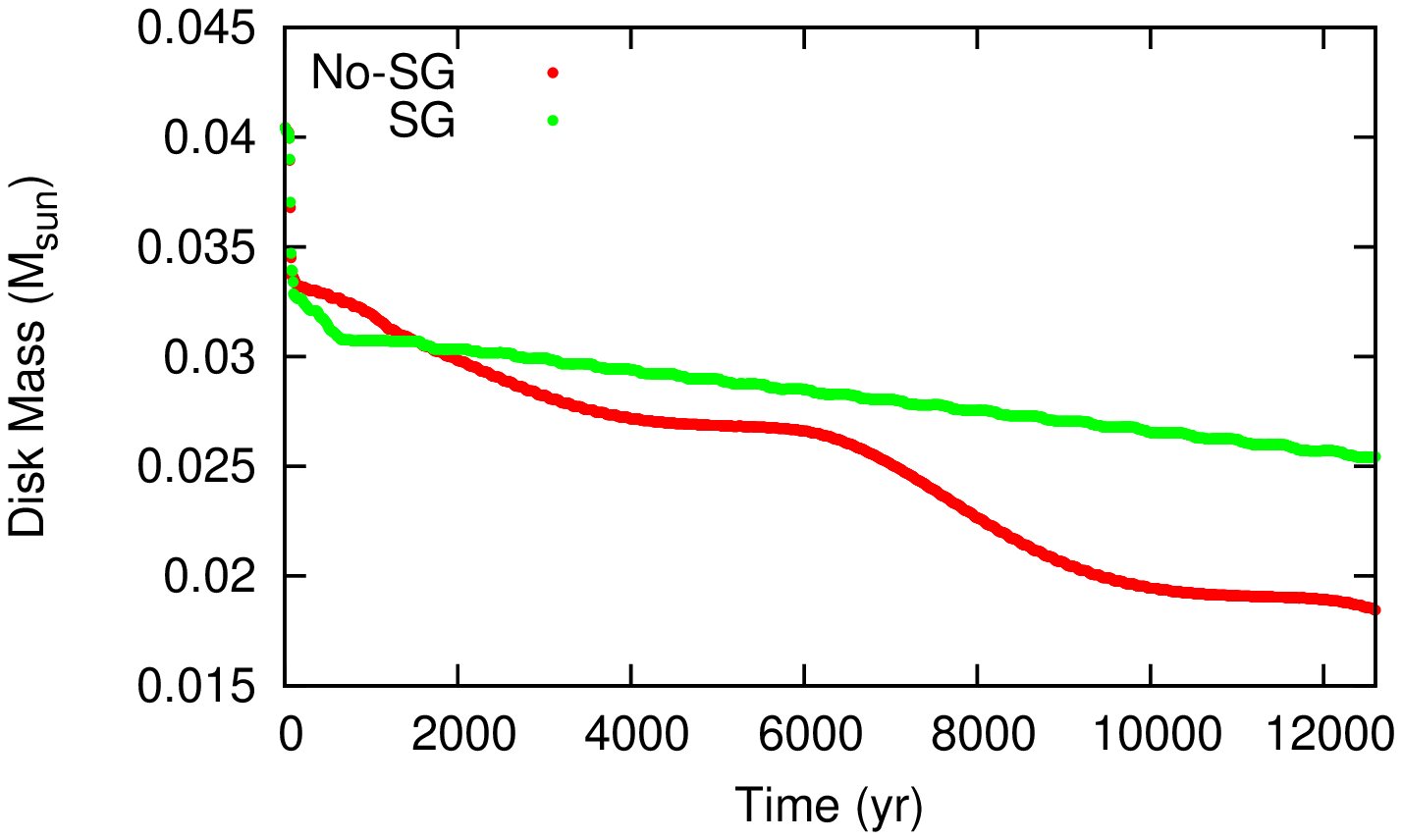}} 
\resizebox{90mm}{!}{\includegraphics[angle=-90]{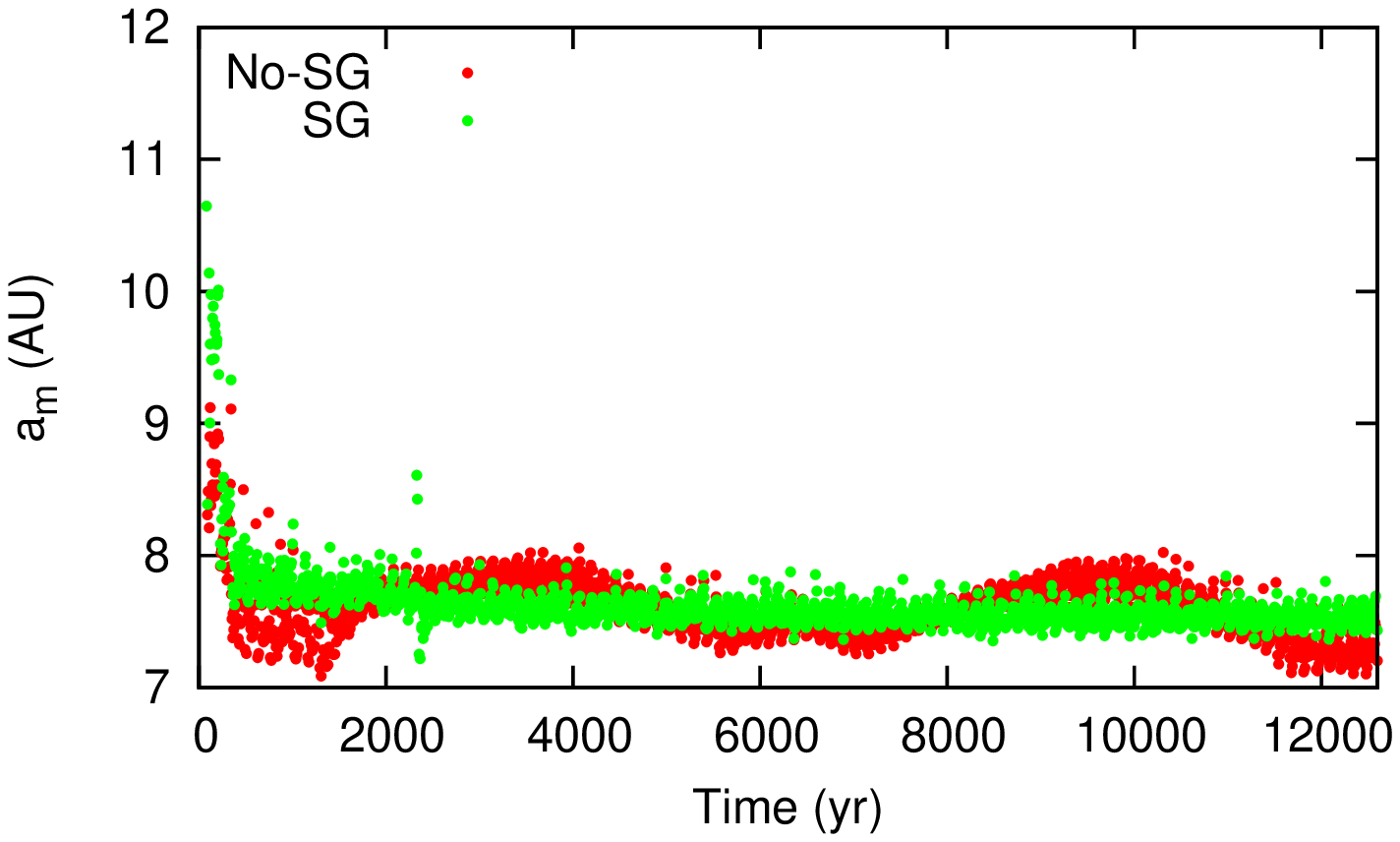}} \\
\end{tabular}
\caption[]{Comparison between the evolution of a disk in a binary 
($a_b = 30$ AU, $e_b = 0.4$) with and without SG (Self Gravity). 
The plot on top to the left (a) shows the variation of $e_d$ with time
while that on the right (b) illustrates its orientation $\varpi_d$.
The middle plots (c,d)  give the corresponding
behaviour of $e_m$ and $\varpi_m$, respectively. 
The bottom plot on the left (e) shows 
the mass of the disk vs. time while that on the 
right (f) illustrates the evolution of the disk 
size described by $a_m$, the effective semimajor axis of 
the elliptical disk.  In all plots the self--gravitating 
case is represented by green dots while that without 
self--gravity by red dots.
}
\label{f2}
\end{center}
\end{figure*}

Fig.\ref{f2} illustrates the effect of SG on the 
evolution of the eccentricity, orientation, size and 
mass loss of the disk over about 90 binary revolutions. 
Both the dynamical and morphological 
eccentricities $e_d$ and $e_m$ are significantly smaller
in the case with SG and almost constant in the considered 
time interval, with $e_d$ being systematically larger than
$e_m$. The case without SG shows large oscillations in 
$e_d$ qualitatively
similar to the behaviour described in \cite{paard08} (excited case) 
even if the disk mass and the binary parameters
of our standard case are different from those adopted
by \cite{paard08}. $e_m$ is lower but still it has a wide 
wavy pattern. Spikes in the disk eccentricity appear whenever
the companion star approaches the periastron and the
disk is more excited.

\begin{figure}[hpt]
\begin{center}
\resizebox{\hsize}{!}{\includegraphics[angle=-90]{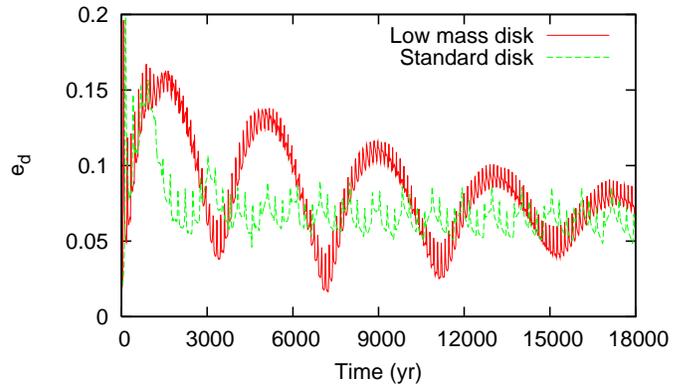}} 
\caption[]{
Variation of the disk eccentricity $e_d$ for a disk with 
initialy 1/10 of the mass of the standard case (0.004 $M_{\odot}$) 
compared to the standard case (0.04 $M_{\odot}$). 
}
\label{lowmass}
\end{center}
\end{figure}

A tentative explanation of why the dynamical and morphological eccentricities of 
a disk are lower with self-gravity is possibly due to the tidal 
interactions between the inner and outer portions of the disk. 
The inner disk acts like a tidal bulge on the outer part 
tending to circularize the orbits of the outer fluid elements
(\cite{murde}). {\bf This triggers a chain effect by which each inner 
ring forces a lower eccentricity on the outer one until a 
steady eccentricity distribution is reached. This stationary distribution 
has a lower eccentricity also in the inner regions, as it can be seen 
from Fig.\ref{tre} (upper panel).}
There is also a delay between the inner bulge and the 
outer one, as it can be seen in Fig.\ref{tre} (middle panel). The perihelia 
of the inner gas cells are different from those of the outer ones 
and the transition is around 3 AU. The inner and outer parts
of the disk exchange angular momentum and energy and, as a 
consequence, there is a tendency towards circularization of 
the orbits. This may lead to an overall lower eccentricity of a
self-gravitating disk under the perturbations of an outer 
star. This explanation related to the tidal bulge appears more 
reasonable by inspecting Fig.\ref{lowmass} where the 
evolution of a lower mass disk is illustrated. In this case 
the initial mass of the disk is 1/10 respect to our standard case
($Q \sim 26$ in the central disk at $t=10^4$ yr). 
The disk eccentricity still approaches a low value of dynamical 
eccentricity very close to that of the standard case also shown in the 
figure as a reference case
but on a longer timescale with large oscillations. Self--gravity acts 
on a longer timescale and the damping is slow, possibly because of 
the lower mass in the tidal bulge,  but it is still effective. We performed also 
three simulations with progressively decreasing intial mass and we find that self--gravity 
is still important when the mass of the disk is 20 times smaller than the 
standard case ($Q \sim 58$) but it does not affect the disk evolution when the
initial mass drops below 50 times that of the standard case
($Q \sim 150$). In
this latest case the behaviour of the disk eccentricity is close to 
that of the standard case without self--gravity in terms of 
disk eccentricity $e_d$. This means that 
indeed the parameter measuring the relevance of self--gravity might be 
$h Q$ as suggested in \cite{paard08}.

An additional effect that might explain why self--gravity keeps the 
disk eccentricity to a lower value is related to the pericenter passage of 
the star. As discussed in Section 5 and shown in 
Fig.\ref{pericenter}, during the close approach of the star to the 
disk at the pericenter strong waves are excited and the 
overall shape of the disk 
becomes highly ellipsoidal. This phenomenon was already observed in 
\cite{kle8}. Self--gravity might damp perturbations on the disk 
as it appears in a different physical scenario concerning
the effects of moonlets on Saturn rings 
\cite{leste}. A faster damping of the pericenter perturbations of 
the companion star possibly leads to a less excited disk. 

Also the evolution of the disk orientation is significantly different 
when self--gravity is turned on. As illustrated in Fig.\ref{f2} top right panel,
$\varpi_d$ 
librates around $\pi$ instead of circulating with a period of 
about 5000 yrs as in the 
case without SG. A similar thing is observed for the morphological 
elements except that $\varpi_m$, in the case with SG, librates around
 $\pi / 3$. We will see in a subsequent section that the center of 
libration for $\varpi_m$ depends on the binary eccentricity. 

The total mass of the disk has a significantly different evolution in the 
case with SG as compared to the case without SG (Fig.\ref{f2}, bottom left panel). 
The mass loss in the case with SG is significantly reduced 
and it is almost linear compared to that without SG. 
This confirms that with SG the disk relaxes
into an almost steady state where its overall 
dynamical properties are almost constant in time
or adiabatically changing on a long timescale. On the other 
hand, without SG the disk appears to be more excited with 
its mass and eccentricity distributions still evolving 
after 12000 yrs. 

The semimajor axis $a_m$ describing the disk shape is 
about 7.8 AU, which is larger than the critical semimajor axis of 6.5 AU
for dynamical
stability computed by \cite{howi}.
However, this is the critical semimajor axis for long term 
stability (1 Myr) of massless particles under the action of 
gravity alone. It is reasonable that a disk, which is also 
affected by pressure forces and viscosity may behave 
slightly differently.
In addition, our simulations last only $1.2 \times 10^4$ yrs 
while the stability limit of \cite{howi} is defined over a 
timescale of 1 Myr.

In conclusion, SG leads to an almost stationary disk on a short timescale
and its internal structure appears to be more compact and 
less eccentric than in the non--SG case. The orientation of the 
disk librates around a fixed value, a behaviour similar to the 
evolution of massive bodies under a perturbative frictional force
(\cite{mascho00}).


\section{Dependence of disk eccentricity on $e_b$}

Possible sources of disk eccentricity in our configuration
are the forced component of eccentricity excited by the 
companion star, 
mean motion resonances between the companion star and disk 
gas particles (which include Lindblad and corotation resonances) and 
viscous overstability (\cite{Kato,laol}). 

We cannot 
attribute the disk eccentricity to the 3:1 mean motion resonance
(inner eccentric Lindblad resonance with m=2),
as described in \cite{lub91},
in all our cases. Our scenario includes a massive 
($M_s = 0.4 M_{\odot}$) and eccentric 
companion star which truncates the disk inside 
the 3:1 mean motion resonance.
As described in \cite{artilub}, the disk truncation moves closer 
to the star for larger values of $e_b$. 
A good measure
of the expected truncation limit is given by the already quoted
2--body stable zone
derived by \cite{howi}. According to their numerical simulations,
the limiting semimajor axis for massless bodies orbiting the primary star
ranges from about 10.5 AU when
$e_b = 0.0$ to 5.7 AU when $e_b = 0.4$ and to 3 AU when  $e_b = 0.6$.
These values reasonably reproduce the size of our truncated 
disks. 
Only in the case with $e_b = 0$ the outer 
edge of the disk may be marginally involved with this resonance. 
However, 
other resonances 
are within the disk and they are given in Table \ref{t:reso} with the
values of the semimajor axes were they are located. 
These values are computed within the 3--body model 
without accounting for pressure forces that, however, do
not dramatically change the location of the resonances, at least
for what it concerns the present reasoning.

\begin{table}[!t]
\begin{center} 
\begin{tabular}{|c|c|c|}
\hline 
\hline
                  i   &     j   &     a (AU)\\
\hline
                  3   &     1   &     12.89\\
                  10  &     3   &     12.01\\
                  7   &     2   &     11.63\\
                  4   &     1   &     10.64\\
                  9   &     2   &      9.84\\
                  5   &     1   &      9.17\\
                  6   &     1   &      8.12\\
                  7   &     1   &      7.32\\
                  8   &     1   &      6.70\\
                  9   &     1   &      6.19\\
\hline 
\hline 
\end{tabular} 
\end{center}
\caption[]
{Location of the mean motion resonances within the circumprimary disk 
up to order 10. The critical argument is of the type $i \lambda_c - 
j \lambda_g - n \varpi_c -m \varpi_g$ where $\lambda_c$ is the true
anomaly of the companion star and $\lambda_g$ that of the gas while 
$\varpi_g$ is the perihelion longitude of the gas. $\varpi_c$ is 
the perihelion longitude of the star which is constant in our simulations. 
}
\label{t:reso} 
\end{table} 

While we may be confident that for low orders $i$ the location is close to the 
effective one, when the value of $k = i-j$ becomes large, the pericenter frequency 
$\dot{\omega}$ may play an active role in moving the location of the
resonance from that we have estimated. 

Concerning the resonance effect, one would expect a reduction of the
disk eccentricity for increasing $e_b$ since the number of resonances 
inside the disk is smaller. On the other hand, for larger values of 
$e_b$ resonances are stronger and, in spite of their large values, 
they might anyway cause some instability in the disk. 

We do not expect a signficant contribution from viscous overstability in
exciting disturbances in the disk since in the standard case we observe
a similar behaviour for different values of viscosity. 

\begin{figure*}
\begin{center}
\resizebox{\hsize}{!}{\includegraphics[angle=-90]{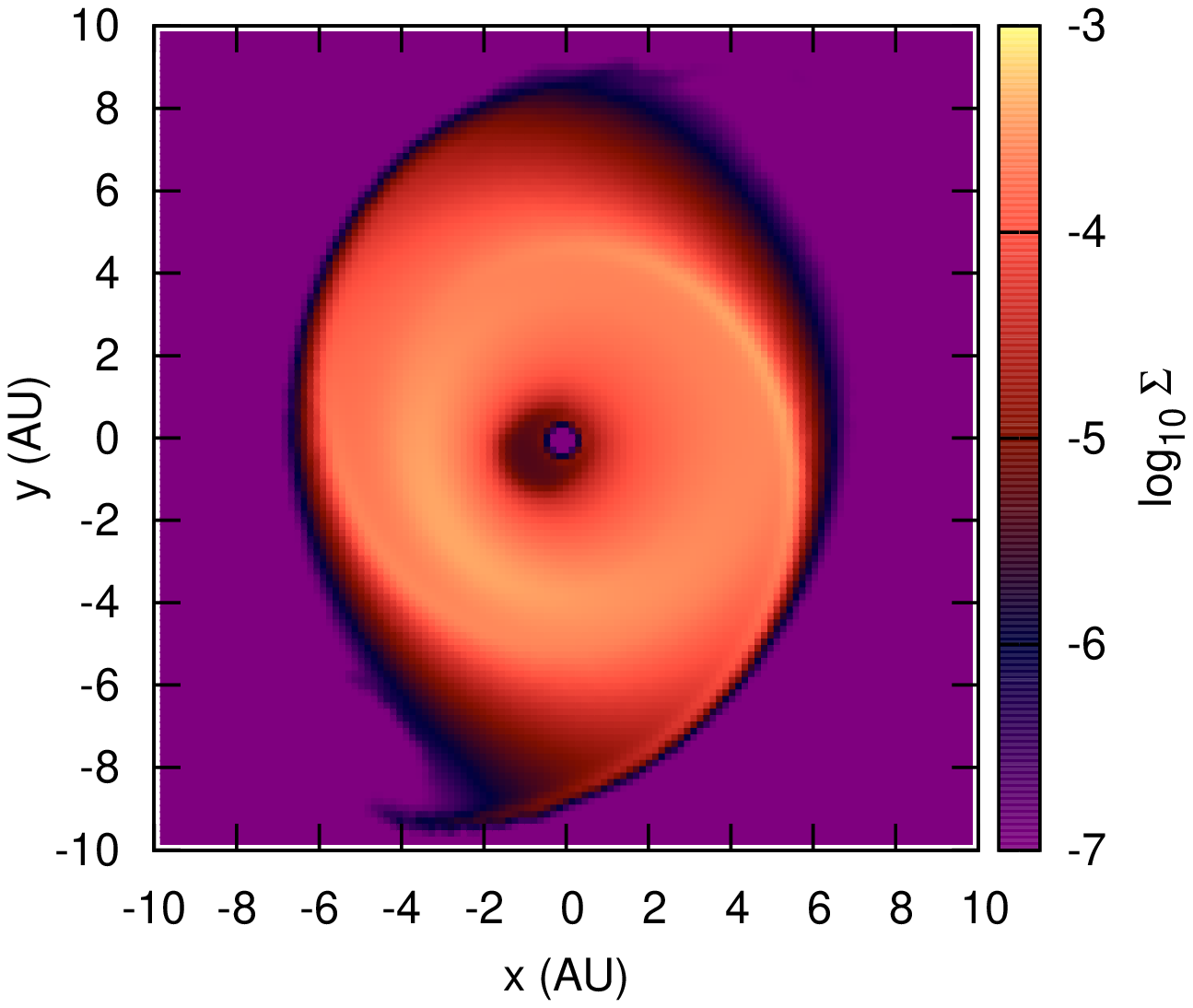}} 
\caption[]{Snapshot of the disk surface density for our 
standard case after 15000 yrs of 
evolution. The companion star is at the close approach 
and it excites strong spiral waves in the disk and it significantly 
alters its shape. 
}
\label{pericenter}
\end{center}
\end{figure*}

An additional effect particularly relevant when the binary 
eccentricity $e_b$ is larger, is the strong gravitational 
perturbations excited by the companion star during the 
pericenter passage. In Fig. \ref{pericenter} we show the 
strongly altered shape of the disk when the companion star
is passing at the pericenter of the binary orbit. This strong
disturbance is slowly damped when the star departs from the 
pericenter. This effects seems somehow to suggest that 
a higher value of $e_b$ favors larger values for both $e_d$ and $e_m$.

On the other hand, a companion star in a very eccentric orbit
spends significantly more time far away from the disk's 
vicinity as compared to the case when
the companion moves on a circular orbit. The effects of 
the strong perturbations during the periastron passage
are damped down on a longer timescale.
As a consequence, it is not obvious
to predict 'a priori' the effects of the pericenter 
passage of the compan ion star on the disk eccentricity 
for different values of 
$e_b$. 

\begin{figure*}
\begin{center}
\begin{tabular}{c c}
\resizebox{90mm}{!}{\includegraphics[angle=-90]{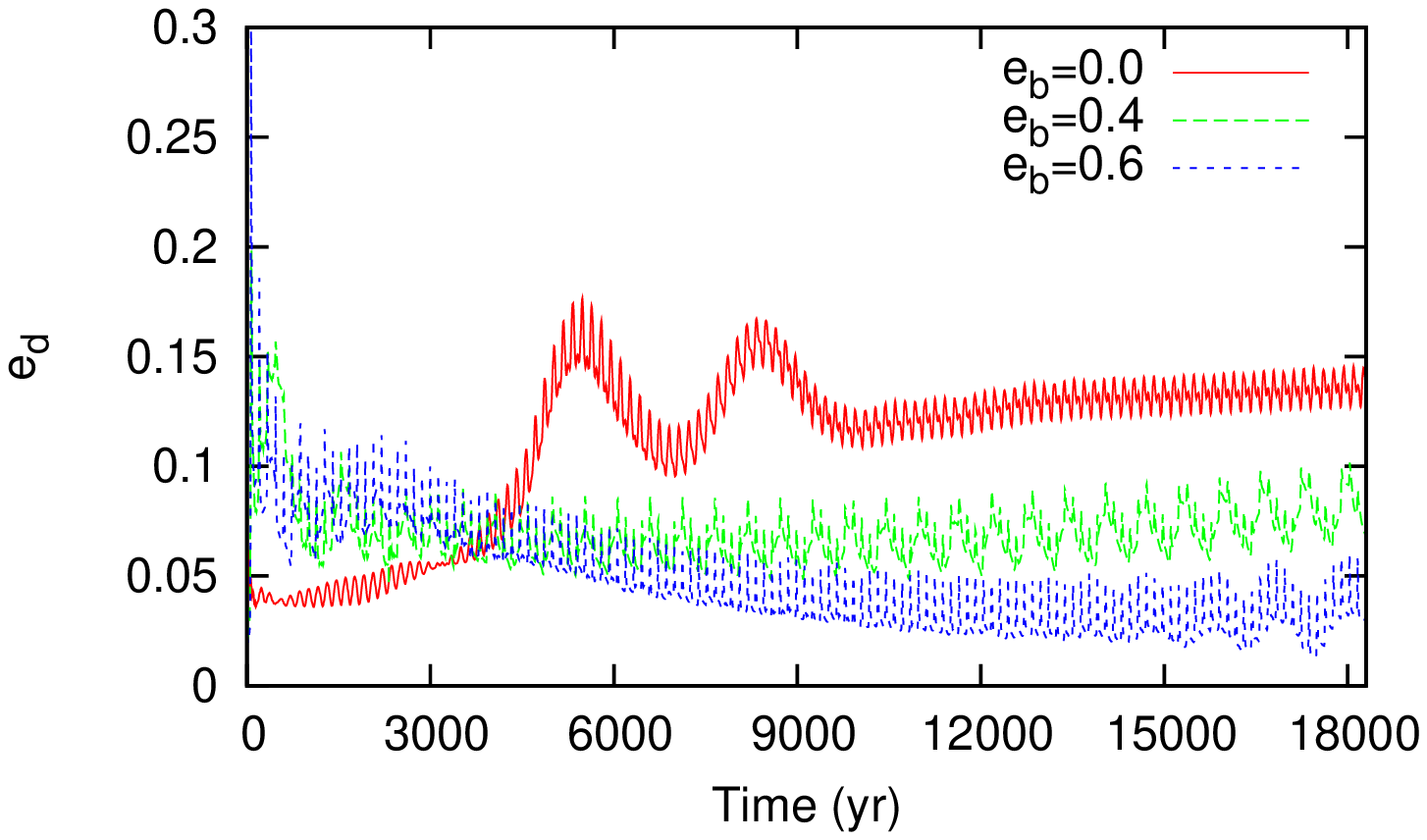}} 
\resizebox{90mm}{!}{\includegraphics[angle=-90]{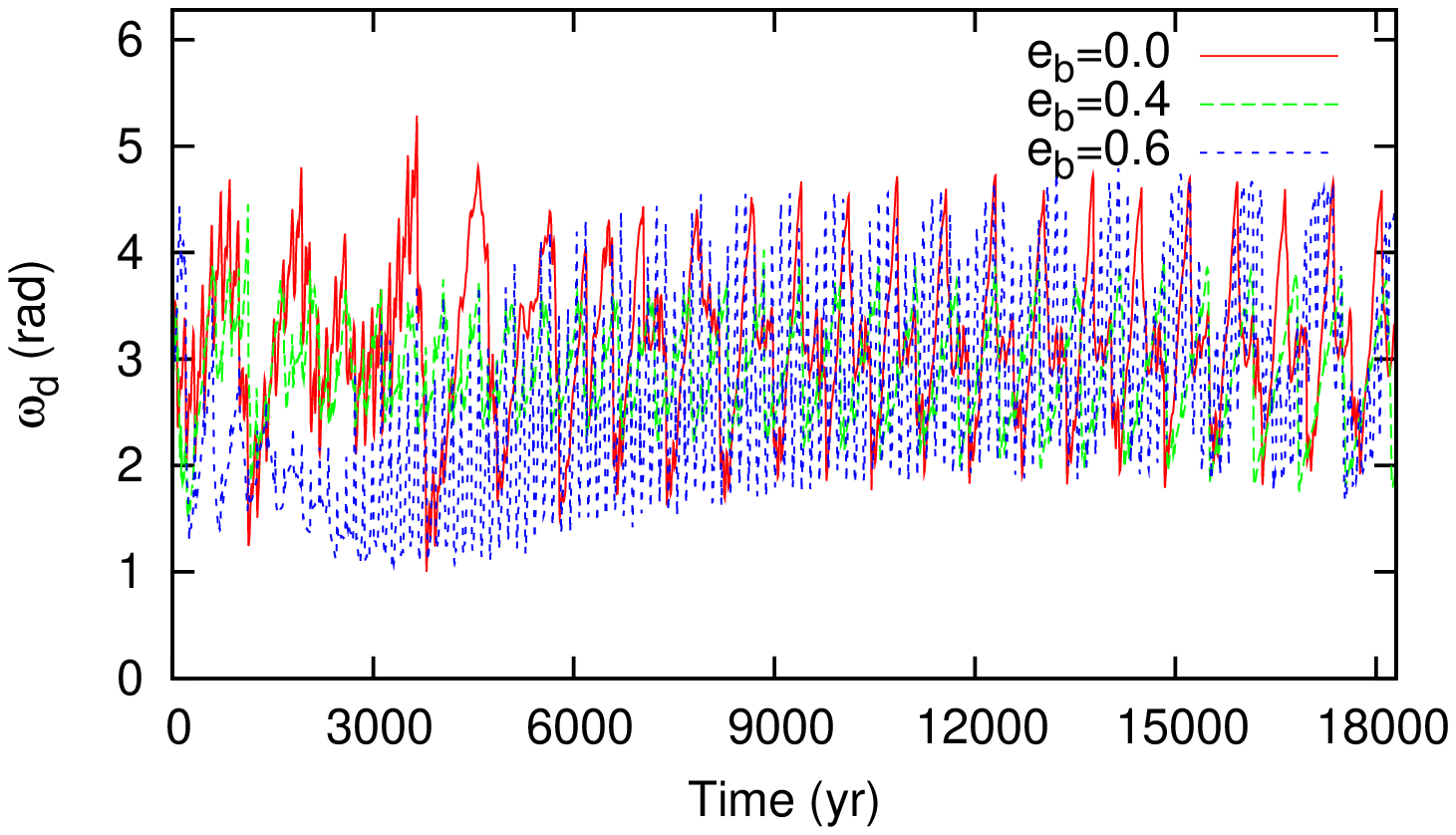}} \\
\resizebox{90mm}{!}{\includegraphics[angle=-90]{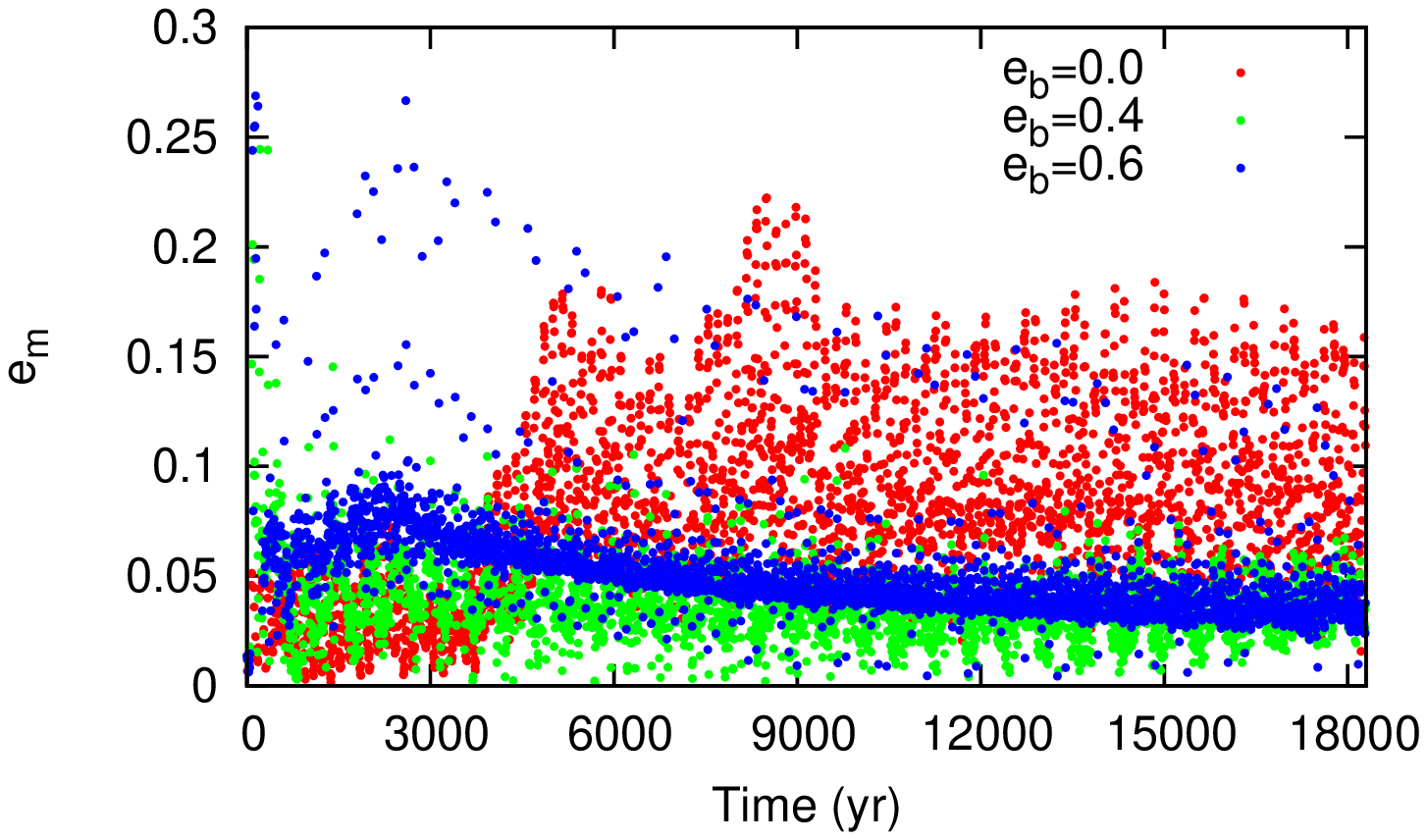}} 
\resizebox{90mm}{!}{\includegraphics[angle=-90]{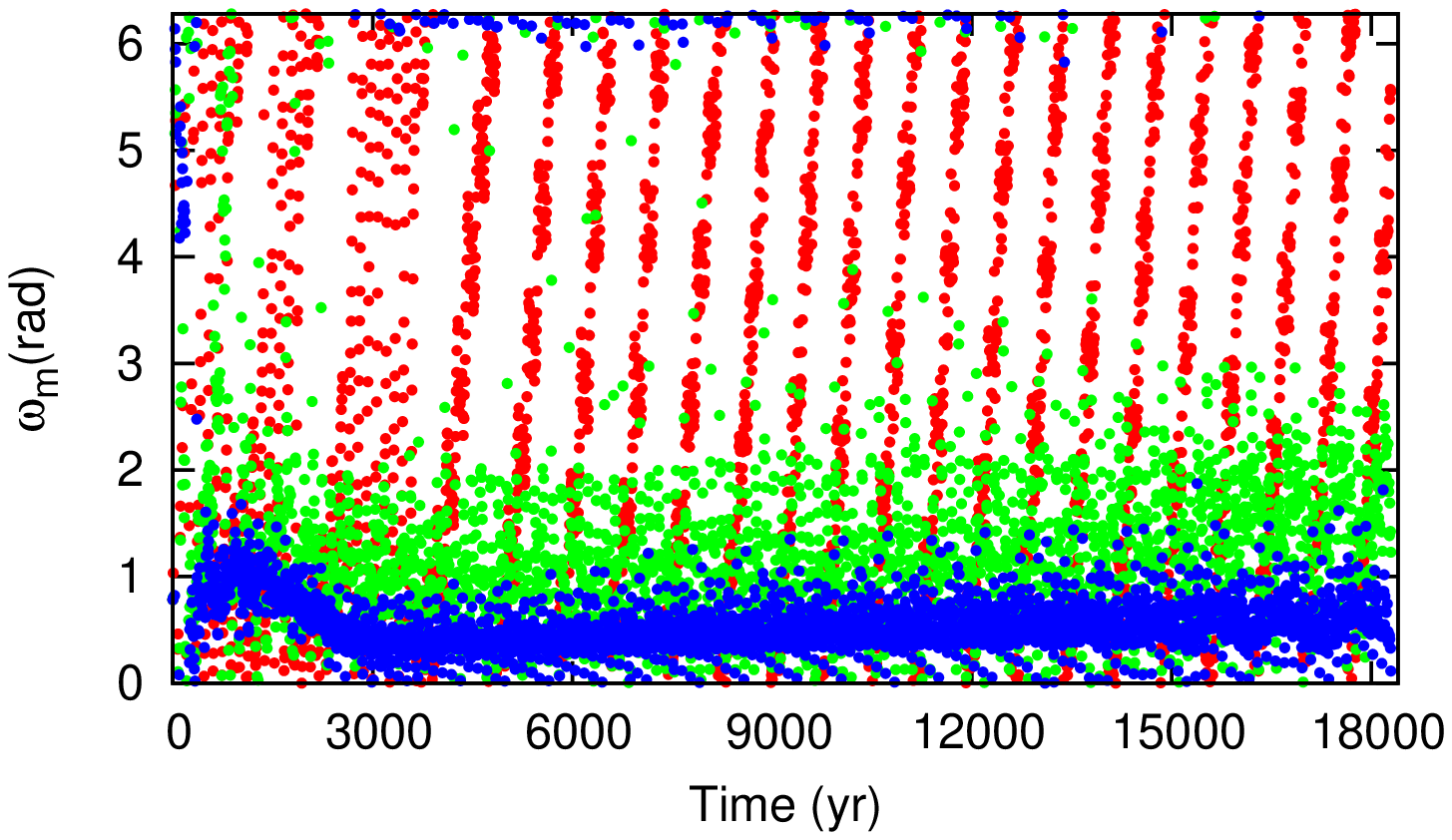}} \\
\resizebox{90mm}{!}{\includegraphics[angle=-90]{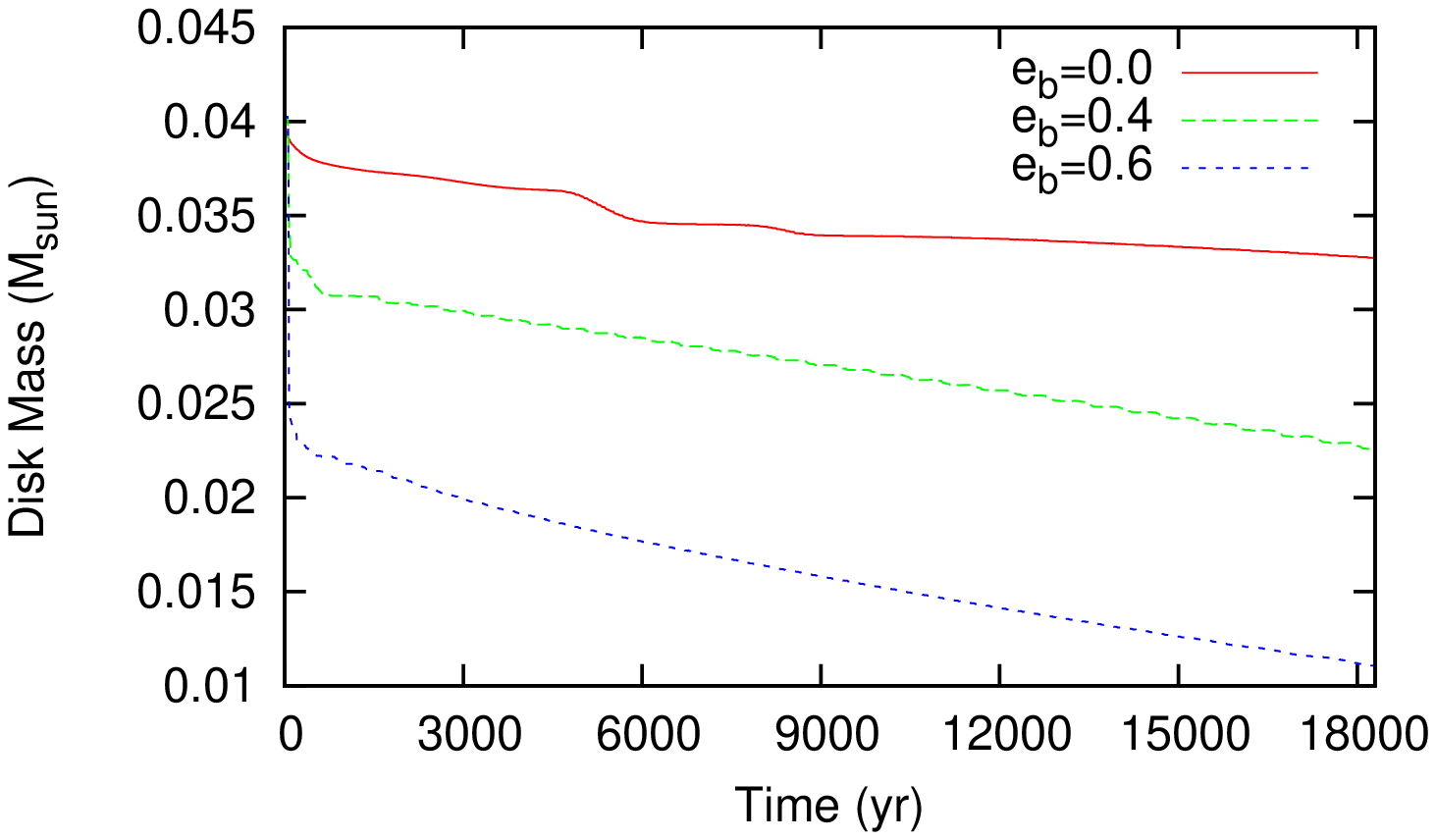}} 
\resizebox{90mm}{!}{\includegraphics[angle=-90]{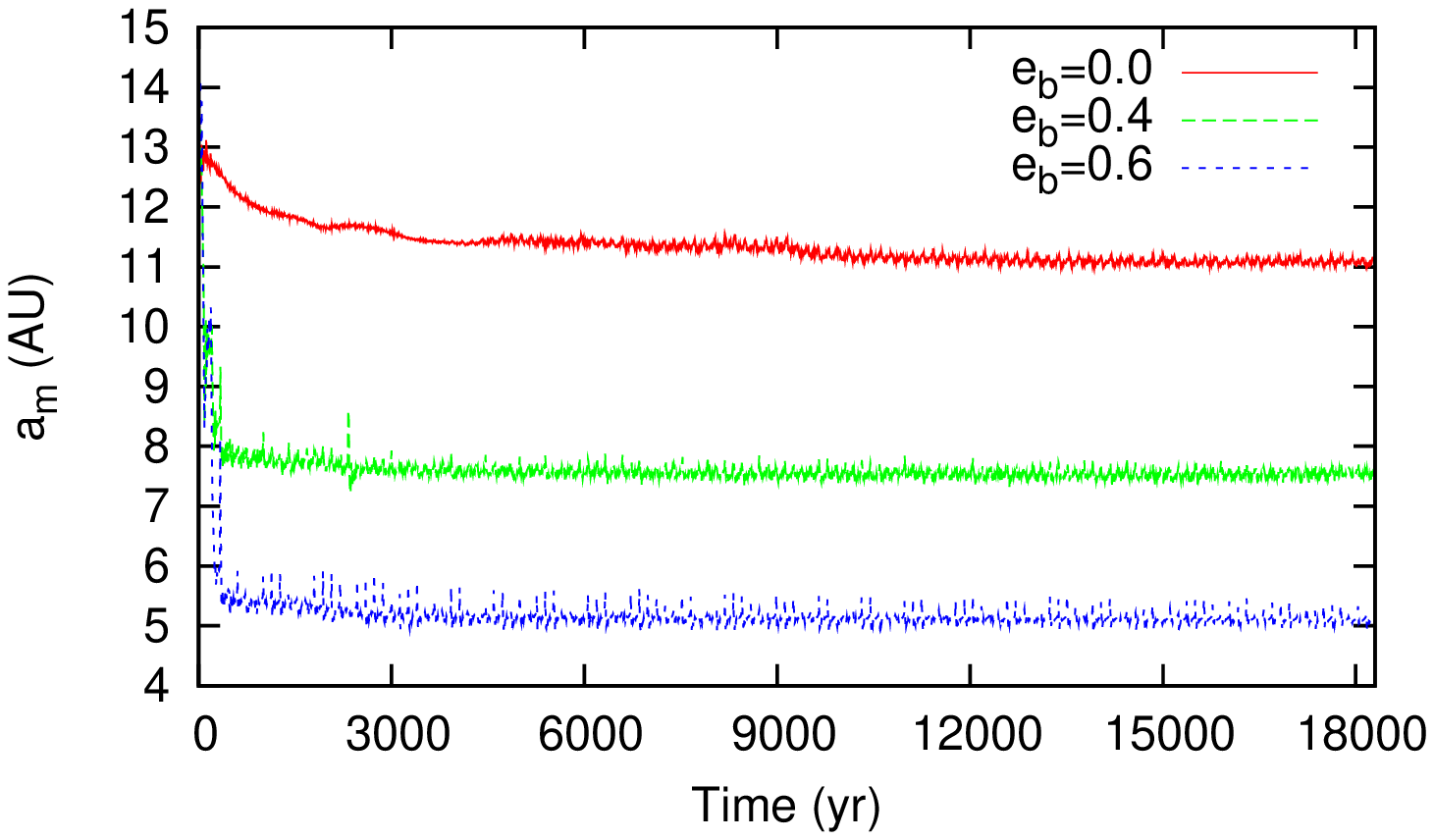}} \\
\end{tabular}
\caption[]{Dynamical and morphological elements for different binary eccentricities 
}
\label{f4}
\end{center}
\end{figure*}

In order to test the dependence of both $e_d$ and $e_m$ on $e_b$ we have
performed 7 long term simulations with values of $e_b$ ranging
from 0.0 to 0.6 with fixed values for $a_b$, viscosity and 
initial disk mass. We included SG in all the models. 
In Fig.\ref{f4} we show 3 selected cases  
with $e_b=0.0$, $e_b=0.4$ (our standard case) and 
$e_b=0.6$.
Both the dynamical 
and morphological eccentricities $e_d$ and $e_m$ indicate that 
larger values of $e_b$ lead, on average, to lower disk eccentricities. 
It implies that the stronger disturbances induced by a more
eccentric companion star
during the fast perihelion passage are not sufficient to stir up 
on average the 
eccentricity of the disk. Eccentric binaries are less effective in 
exciting the disk eccentricity. 

The effect of $e_b$ on $\varpi_d$ is not significant and 
$\varpi_d$ librates around $\pi$ for any value of $e_b$. Analyzing
the evolution of $\varpi_m$, however, we notice that its behaviour 
strongly depends on $e_b$. For $e_b = 0.0$ and $e_b = 0.1$ the orientation of the disk 
described by $\varpi_m$ circulates with a period of some hundred 
years, while for larger binary eccentricities it librates around a 
decreasing value. For $e_b = 0.4$ it librates around $\pi/2$ while for 
$e_b = 0.6$ it librates approximately around $\pi/3$. 
Self--gravity forces the disk appear to behave similarly to a small body 
under the action of a dissipative force. As shown in \cite{mascho00},
an increasing eccentricity of the perturber causes the periastron 
of a small planetesimal orbit, which is perturbed by gas drag, to be aligned 
with increasingly smaller values of $\varpi$. 

The disk eccentricity measured by the morphological $e_m$ shows a
behaviour similar to $e_d$ for different values of $e_b$. However, 
$e_m$ is always lower on average than $e_d$. The case with 
$e_b = 0.0$ shows large oscillations around 0.1 with the same 
period of circulation of  
$\varpi_m$. For the case with $e_b = 0.6$, points well above the 
average are observed. They are computed by the algorithm estimating
$e_m$ when the disk is strongly perturbed in correspondence to the periastron
passage of the companion star and they are not significant. 

The semimajor axis of the disk estimated by $a_m$ is decreasing as
a function of $e_b$, as expected. 
We recall here that the limiting semimajor axis for stability 
derived by \cite{howi} is smaller when compared to $a_m$ 
(see  Fig.\ref{f4} bottom left panel)
for any value of $e_b$. The timescale
we are considering here ($1.8 \times 10^4$ yr, i.e. 130 binary 
revolutions) is possibly not long enough to fully destabilize the 
outer border of the disk. This is confirmed also by the time
evolution of the disk mass. We are not expecting that 
the disk border relaxes exactly at the limiting semimajor axis 
given by \cite{howi} because of the presence of pressure forces 
and viscosity which alter the dynamics of individual gas molecules 
from a pure 3--body problem. However, it is a good reference value 
for the size of the disk.  

In Fig.\ref{f4} bottom right panel,  we illustrate the progressive mass loss of the 
disk. After the initial large shrinking of the disk because of 
the tidal truncation, the mass loss continues at different rates
depending on $e_b$. The responsible mechanisms are:

\begin{itemize}

\item mass in--fall of disk material on the star caused by the 
viscous evolution.  

\item mass stripping during the perihelion passage of the 
companion star

\item progressive dynamical destabilization of the outer parts of the disk 
\end{itemize}

\begin{figure}[hpt]
\begin{center}
\resizebox{90mm}{!}{\includegraphics[angle=-90]{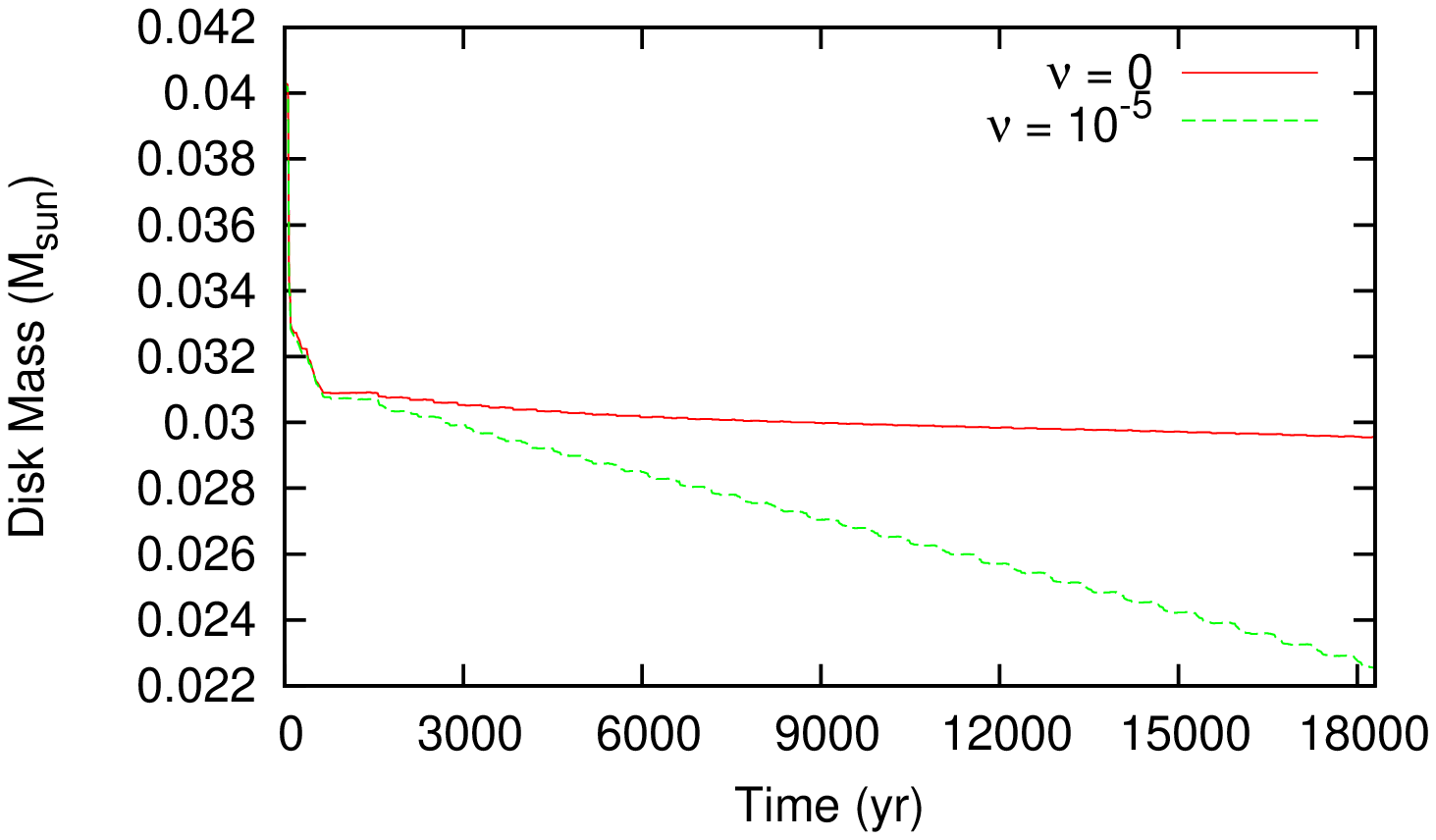}} 
\resizebox{90mm}{!}{\includegraphics[angle=-90]{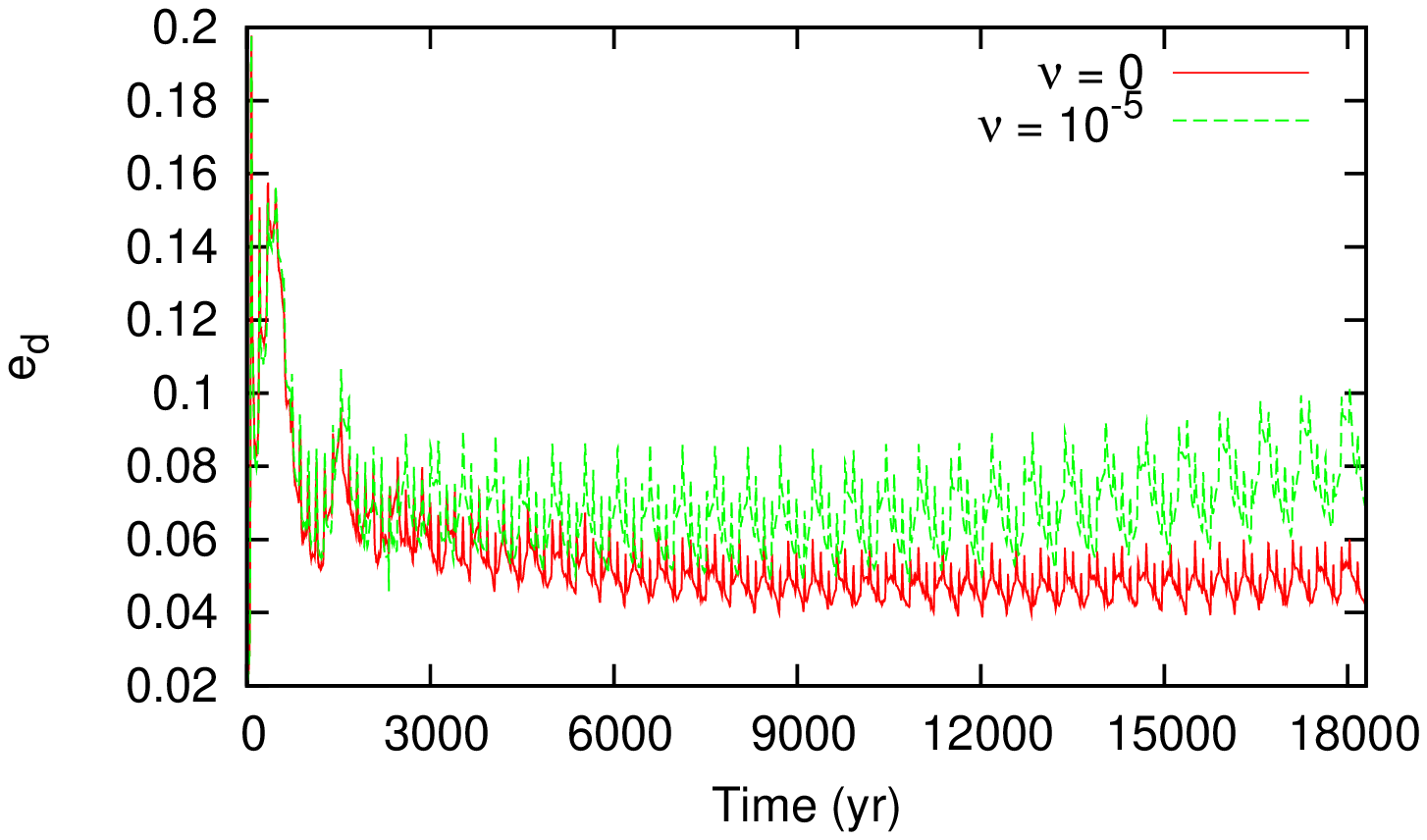}} 
\caption[]{Evolution of the disk mass and eccentricity
($e_d$) for the viscous and inviscid case.
}
\label{f5}
\end{center}
\end{figure}

The most effective mechanism appears to be the viscous evolution. 
In Fig.\ref{f5} top left panel we show the mass loss as a function of time for 
our standard case and the corresponding inviscid case. 
The inviscid model has a comparatively much slower mass loss which may 
be ascribed to the other two above mentioned mechanisms. 
In addition, when the companion star passes through the pericenter, 
the eccentricity of the disk is temporarily excited and this may cause 
a further in--fall of material across the inner border because the gas
particle perihelion gets closer to the star. 

\begin{figure}[hpt]
\begin{center}
\resizebox{90mm}{!}{\includegraphics[angle=-90]{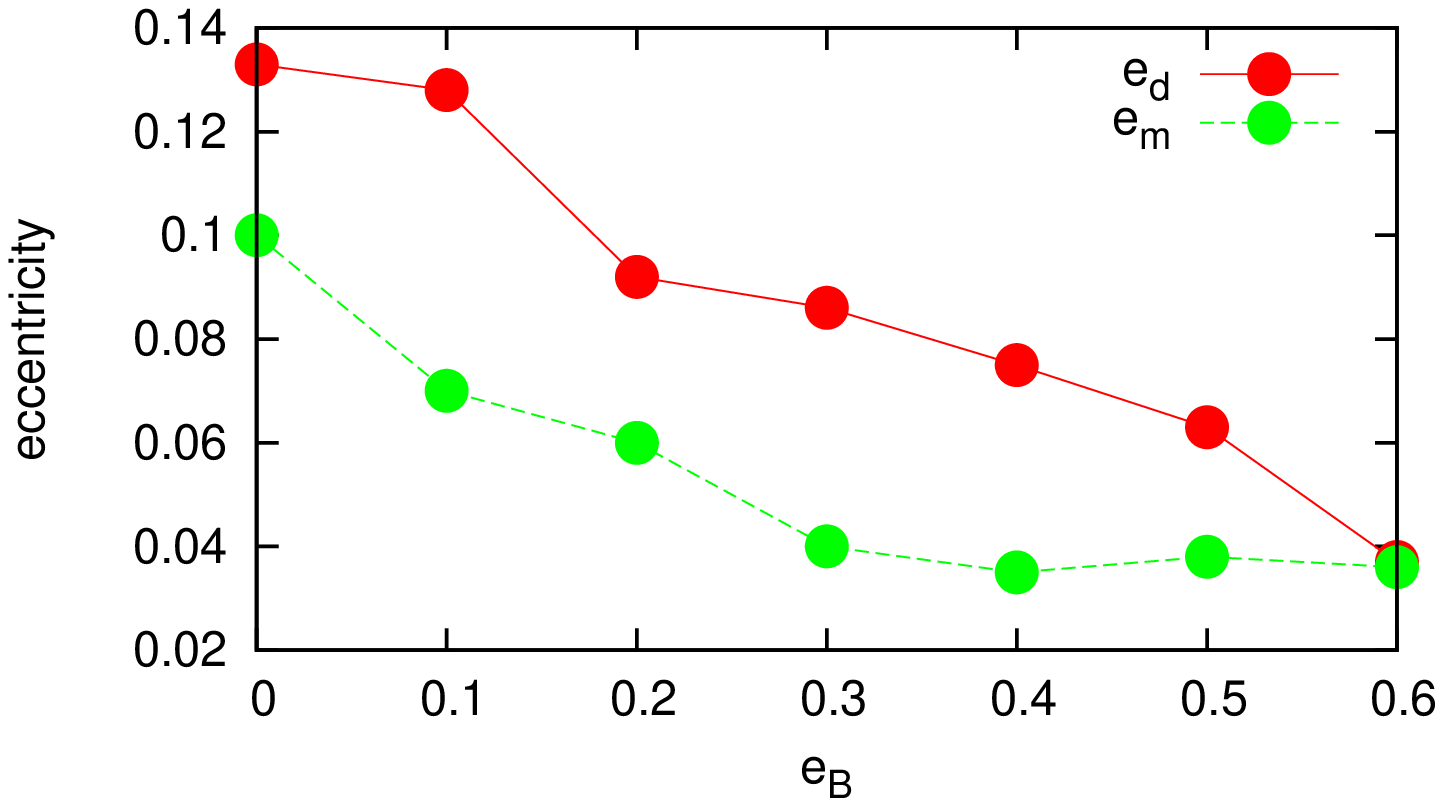}} 
\resizebox{90mm}{!}{\includegraphics[angle=-90]{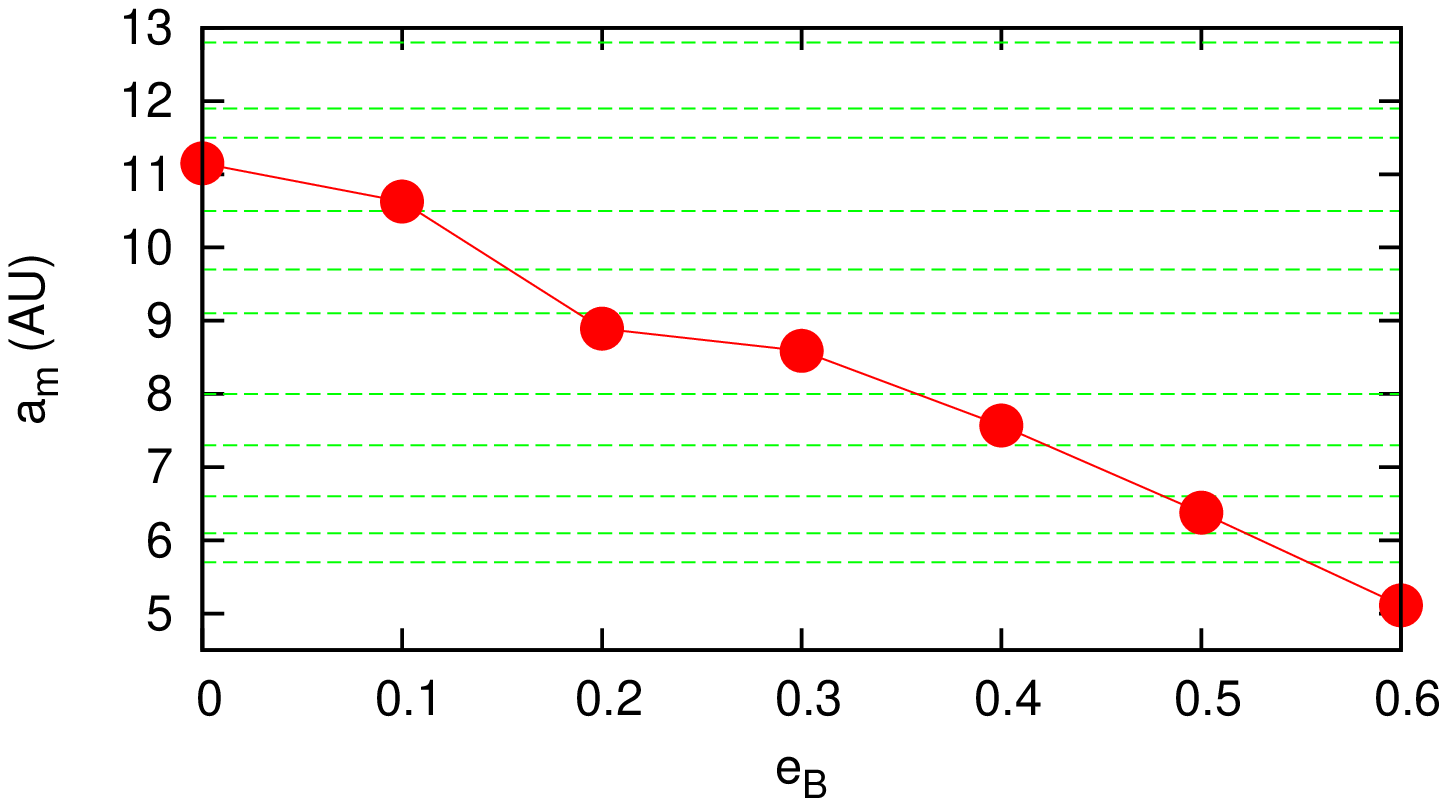}} 
\caption[]{Variation of the disk eccentricity (upper plot), measured by $e_d$ and
$e_m$, and of the disk size given by $a_m$ (lower plot) 
for different values of the binary eccentricity $e_b$. The dashed lines 
in the lower plot mark the location of the mean motion resonances
with the companion star up to order 9 and degree 10. 
}
\label{f6}
\end{center}
\end{figure}

The dependence of the disk eccentricity and size on the binary 
eccentricity is summarized 
in Fig.\ref{f6} for all our simulations. Both $e_d$ and $e_m$ show a 
decreasing trend for increasing $e_b$ confirming that the perturbations
are stronger in the circular case where a tidal response is constantly 
forced. 
It is noteworthy that this behaviour is 
not typical of self-graviting disks only. 
We performed 3 simulations with 
$e_b = 0.0, 0.4$ and $0.6$, respectively, excluding self--gravity
from the model.
The dynamical eccentricity
$e_d$ was decreasing from an average value of 0.42 for $e_b = 0.0$ to 
0.25 for $e_b = 0.4$ and  0.2 for $e_b = 0.4$.
The size of the disk measured 
by $a_m$ shows an almost linear reduction for larger values 
of $e_b$. This is due to the increasing strength of higher order 
mean motion resonances of disk particles and the companion star. 
In Fig.\ref{f6} bottom panel, together with $a_m$ the location of 
all resonances up to order 10 are illustrated 
by dashed lines. The different truncation
radius related to $e_b$ also changes the response of the disk to eccentricity
forcing of the companion. Although a larger  $e_b$
leads to stronger forcing, this may be more than compensated for by
the disk truncation radius being closer to the primary.

\section{Discussion and conclusions}

We have analysed the evolution of disks in binary star systems including self--gravity.
 We study 
their properties by using two different set of parameters. The first set 
has been widely used and computes the eccentricity and orientation of the 
disk as a mass weighted average over the disk of the individual 
cell orbital parameters. This method outlines the dynamical properties 
of the disk and is termed by subscript $d$.
The second set is more oriented to observers and 
it measures the morphological ellipticity and orientation of the disk.  
It is termed with subscript $m$ and it is computed by numerically
fitting the outer edge of the disk. 

An important ingredient in modelling the disk evolution is self--gravity. 
We show that it forces the disk to behave more like a solid body in terms 
of orientation of the disk. Instead of circulating, the disk's apsidal line 
librates around a fixed value  in eccentric binary systems  like a 
body moving under the gravitational pull of the two stars and a dissipative 
force. In addition, self--gravity does not allow the disk to become 
very eccentric. 
For a binary eccentricity of $e_b = 0.4$ both the dynamical 
and morphological eccentricity $e_d$ and $e_m$ of the disk are lower than 0.1. 
This is an important result in terms of planetesimal accretion since, as
shown by \cite{paard08}, lower values
of the disk eccentricity give lower impact velocities
than with an eccentric disk. This might lead to
an environment which is less hostile to accretion,
although, even in low $e_d$ cases, the relative impact velocity might be
too high to allow protoplanet formation. 

A somewhat unexpected result is the increase of disk eccentricity
with decreasing binary eccentricity, when keeping the same semimajor 
axis. 
The case in which the binaries are on circular orbits appears
to be the most perturbing configuration in terms of disk eccentricity
and this outcome is typical also of non--self--gravitating disks. 
It is significantly larger than for $e_b = 0.4$ and $e_b=0.6$. 
This is an indication that the strong perturbations 
during the short timespan in which  
the companion star is at the pericenter are damped down when 
the companion is far away. The circular case is more effective 
in exciting a tidal response from the disk.  
This does not necessarily mean that highly eccentric binaries have
a higher probability of forming planets. In fact, the direct gravitational 
perturbations of an eccentric companion on the planetesimals lead 
to larger relative velocities that may not be necessarily damped 
by a low eccentric disk. The outcome in this case strongly depends 
on the interplay between the drag force due to the disk and the 
forced component of the orbital eccentricity of planetesimals 
due to the secondary star (\cite{theb06}). These results are a first step
towards a more comprehensive study made with a hybrid code where the 
evolution of the gaseous component of the disk is computed with a hydrodynamical 
scheme including self--gravity 
while the planetesimal trajectories are computed with a N--Body
integrator.

The evolution of a circumstellar disk in eccentric binaries must be 
further explored by changing additional parameters like the mass 
and semimajor axis of the companion star. 
Also numerical issues should be investigated like the dependence of the 
results on the boundary conditions, even if the one we have adopted 
is widely used. The region near the inner edge of the disk, where the mass flows
inwards, must be explored 
in more detail. In models with higher binary eccentricity, 
we obtain elliptically shaped empty regions which needs confirmation
by more extended simulations which are out of the scope of this paper. 

\begin{acknowledgements} 
We thank an anoymous referee for his comments and suggestions 
which helped to significantly improve the paper.
Computations were performed on the "Mesocentre SIGAMM" machine,
hosted by the Observatoire de la C\^ote d'Azur.
\end{acknowledgements}

\begin{appendix}
\section{Origin of the internal elliptic hole of the disk}

When the gas particles in a disk are on elliptic orbits, imposing a
circular inner edge to the disk  may lead to the formation of
a central elliptic zone of low density. This is an
effect related to the
Keplerian nature of the orbits. When the particles are at
pericenter, they may pass through the inner edge of the
grid and then they are lost. This is simulated in Fig. A.1 left panel
where we compute
$10^5$ two--dimensional elliptical orbits (which can be
associated to gas particles) with semimajor axis
ranging from 0.1 to 3 AU. The eccentricity and pericenter longitude
have similar values to those
computed by the hydrodynamical code ($e_i$, $\varpi_i$)
while the mean anomaly is
selected randomly. During the orbit generation,
anytime the pericenter is lower than 0.5
AU (the inner limit of the grid we have used in the hydrodynamical
simulations) the particle is thrown away. The number density
of the surviving particles is illustrated in Fig. A.1 left panel and it is
very similar to the outcome of FARGO shown in  Fig. A.1 right panel.
Not only the shape of the elliptical low density region is
similar, but there is also in both figures an over--density
located beyond the apocenter of the elliptical shape.

\begin{figure}[hpt]
\begin{center}
\begin{tabular}{c|c}
\hskip -1.9 truecm
\resizebox{80mm}{!}{\includegraphics[angle=-90]{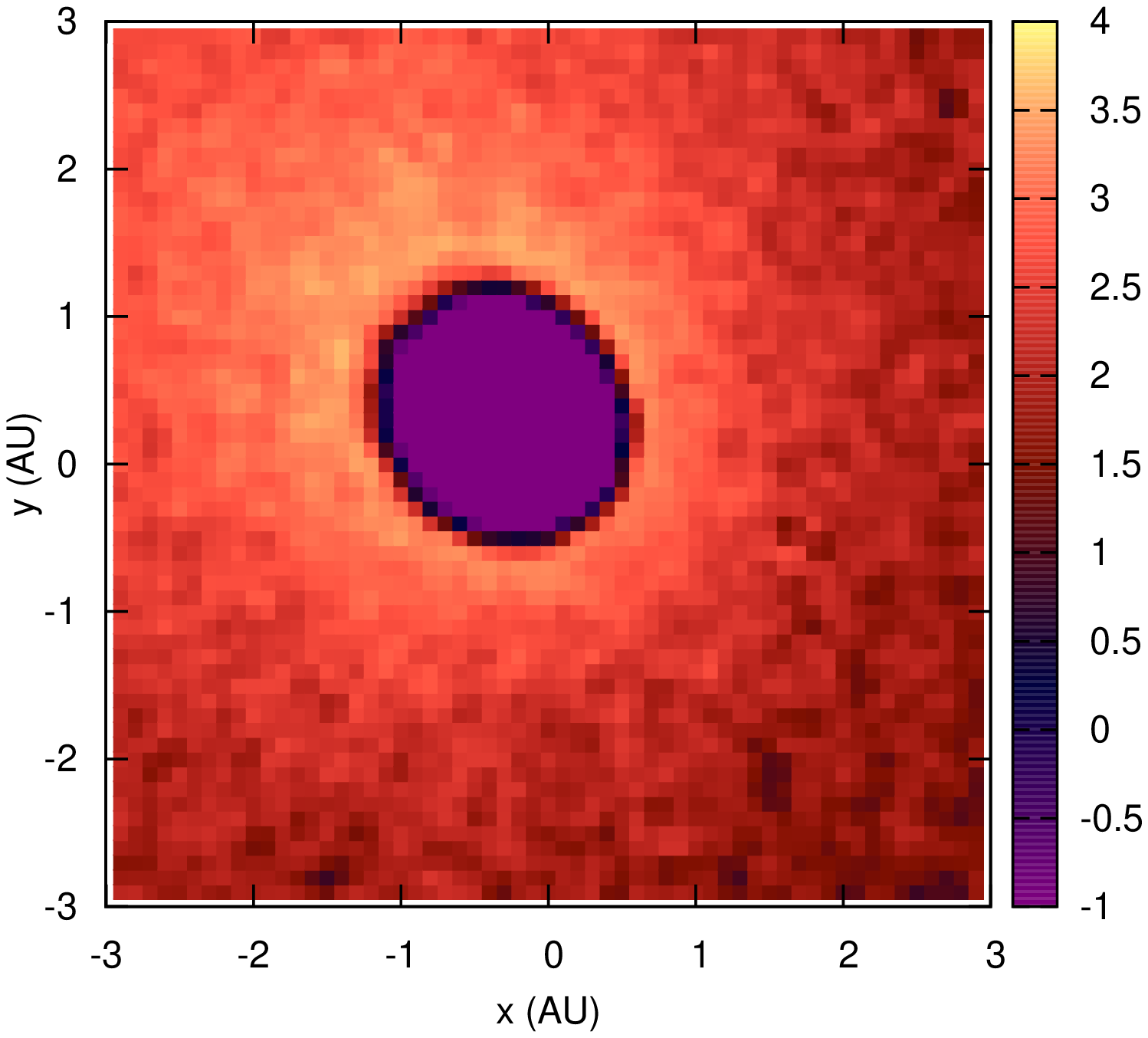}} \hskip -3.3 truecm
\resizebox{80mm}{!}{\includegraphics[angle=-90]{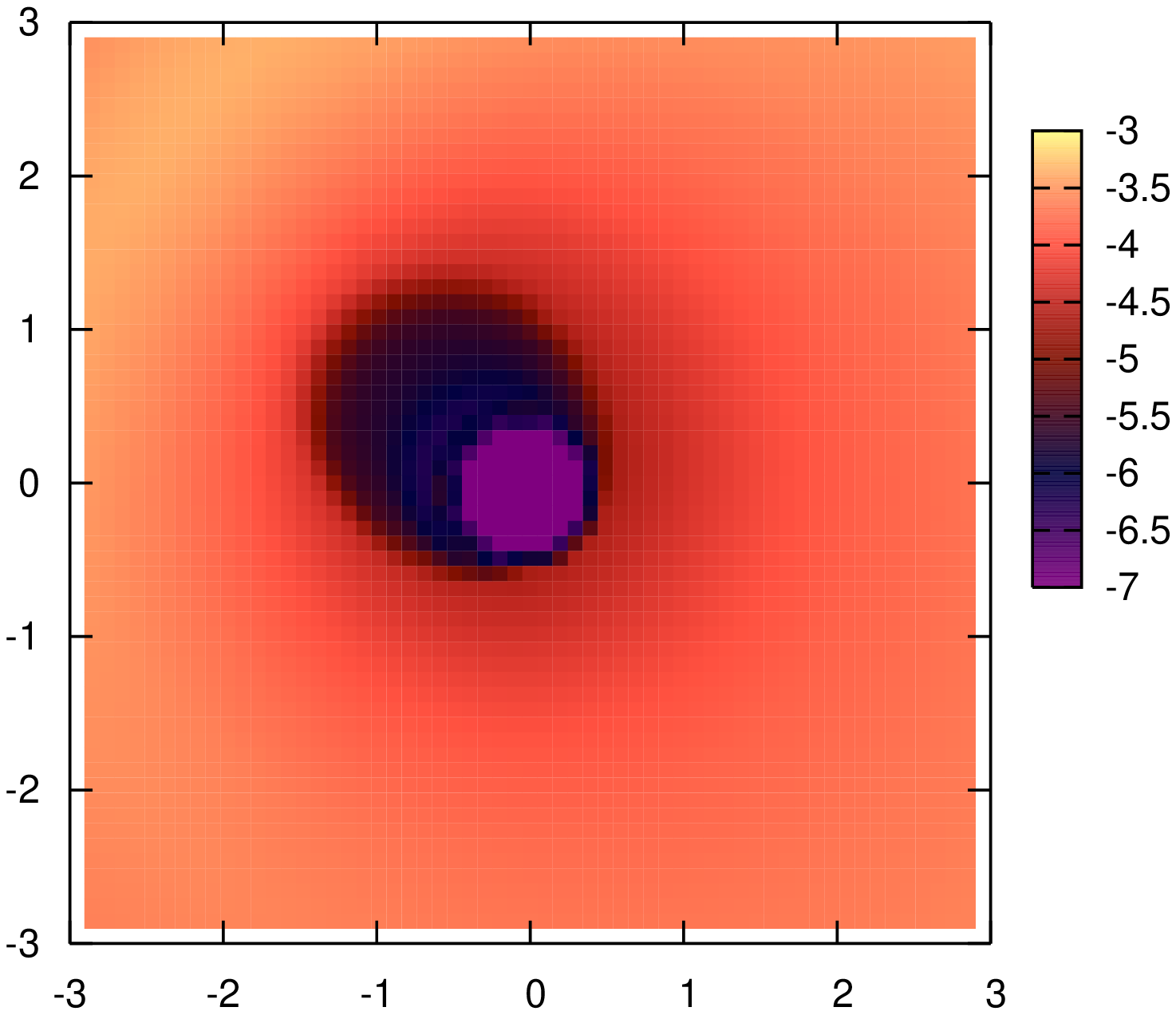}}
\end{tabular}
\caption[]{Number density of a set of test particles in Keplerian
orbits depleted of all those having pericenter lower than
0.5 AU (left plot) and outcome of FARGO (right plot).
The color density in the right plot gives the logarithm
of the gas density in normalized units.
}
\label{f7}
\end{center}
\end{figure}

\end{appendix}

\end{document}